\documentclass[proc]{edpsmath}

\usepackage[utf8]{inputenc}
\usepackage[T1]{fontenc}
\usepackage{graphicx}
\usepackage{todonotes}
\usepackage{amsfonts}
\usepackage{amsmath}
\usepackage{listings}
\usepackage{caption}
\usepackage{subcaption}
\usepackage{hyperref}
\usepackage{multirow}

\usepackage[linesnumbered,ruled,vlined]{algorithm2e}

\begin{document}

\title{Reducing the memory usage of Lattice-Boltzmann schemes with a DWT-based compression}
\author{Clément Flint}\address{clement.flint@inria.fr}
\author{Philippe Helluy}\address{philippe.helluy@unistra.fr}

\begin{abstract}
    This paper presents a new solution to address the challenge of increasing memory usage in high-performance computing simulations of Lattice-Bolzmann or Finite-Volume schemes.
    Our approach utilises a lossy compression scheme based on the Discrete Wavelet Transform (DWT) to achieve high compression ratios while preserving the accuracy of the simulation.
    Our evaluation on two different FV/LBM schemes demonstrates that the approach can reduce memory usage by several orders of magnitude.
\end{abstract}

\begin{resume}
    Ce papier présente une nouvelle solution pour faire face à l'augmentation de l'utilisation de la mémoire dans les simulations haute performance basées sur les méthodes Lattice-Bolzmann ou Volumes Finis.
    Notre approche utilise un schéma de compression avec perte basé sur la transformée par ondelettes discrète (DWT) pour obtenir des taux de compression élevés tout en préservant la précision de la simulation.
    Notre évaluation sur deux différents schémas VF/LBM démontre que l'approche peut réduire l'utilisation de la mémoire de plusieurs ordres de grandeur.
\end{resume}

\maketitle

\tableofcontents

\section*{Introduction}

The Lattice-Boltzmann Method (LBM) and the Finite Volume method (FV) are numerical methods for solving partial differential equations.
These methods are typically used for their ability to accurately simulate fluid flows, for instance.
We refer the reader to \cite{succi2001lattice} for a general description of the Lattice-Boltzmann method and to \cite{eymard2000finite,leveque2002finite} for general descriptions of the Finite Volume method.
In our work, we assume that the discretisation is done using a regular grid.
This means that the simulation domain is divided into a regular 2D or 3D grid of cells.
When the space is discretised on a regular grid, LBM/FV schemes are similar to stencil algorithms.
A stencil algorithm is a type of algorithm where the next state of each cell is a function $\varphi$ of the values of its $N$ relative neighbours.
The arrangement of the $N$ neighbours can be arbitrary. This arrangement is referred to as the stencil of the algorithm.

Our primary concern is to improve LBM/FV simulations that run on a single-GPU machine.
This assumption allows us to have a case where memory usage is by far the most limiting factor of the algorithm, making it a good candidate for the use of data compression techniques.

The use of data compression in the field of high-performance computing has already been explored in several works \cite{astsatryan2020performance,low_latency_LZ4_FPGA}.
Besides memory saving, the use of data compression can also be motivated by the need to transfer data between the CPU and the GPU efficiently.
It is possible to find situations where the cost of the data compression/decompression is largely compensated by the acceleration of the data transfers.
In practice, the compression is achieved through the use of general data compression libraries such as nvCOMP \cite{nvcomp_2020} or zfp \cite{lindstrom2014fixed}.

However, the data handled by LBM or FV schemes have special structures that can be exploited to design a superior compression algorithm.
In many physical applications, the computed fields present large regions with smooth variations, separated by sharp discontinuities.
This is the case, for instance, in the simulation of fluid flows.
In this paper, we design a compression algorithm for FV/LBM schemes that is based on the Discrete Wavelet Transform (DWT), which is well adapted to this type of data.

In the FV framework, multiresolution schemes based on wavelet analysis have a long history. We can mention, among many others,
\cite{harten1996multiresolution, cohen2003fully,abgrall1998multiresolution}. In these works, the general strategy is to represent the data in the multiscale (wavelet) basis and compute directly the evolution of the solution in this basis, i.e. on the locally refined grid.
This leads to very complicated software implementations, with highly non-uniform data access.
The programming is even more complex when parallel and hybrid CPU/GPU computing is addressed \cite{brix2009parallelisation,BursteddeCalhounMandliEtAl14}. 
In the LBM framework, similar works and strategies have been proposed. See, for instance, \cite{bellotti2022multiresolution,yu2009interaction,latt2021palabos}.

In addition to the high complexity, in many practical cases, these implementations are not very efficient.
Finally, they often require a complete rewriting of the FV or the LBM scheme they are based on. 

In this work, we propose and evaluate a much simpler strategy, which avoids algorithms that are difficult to parallelise.
The simple idea is to perform a patch decomposition of the computational domain and apply wavelet compression/decompression in each patch, in turn.

In this way, most of the patches are stored in memory in a compressed form, while the patch that is currently being processed is stored in memory in an uncompressed form.
This allows us to reduce the memory usage of the algorithm.

There are two main difficulties in this approach.
The first point is that the compression is lossy.
One has to verify that this loss is small and can be controlled by the user through a threshold value.
Secondly, when shock waves are present, it is important to build a globally conservative scheme, in order to ensure the convergence of the scheme under mesh refinement \cite{hou1994nonconservative}.
In this work, we propose a simple way to ensure the mass conservation of the wavelet compression algorithm at the boundaries between the patches.

Then, our approach can be used with any standard LBM/FV scheme that uses a regular grid.

In this work, we evaluate the performance of our approach on a single patch.
We show that using this algorithm reduces the memory usage of the tested scheme while preserving its quality.
We also evaluate the cost of the data compression.
We show that the compression time can be significantly lower than the computation of the LBM/FV scheme.
This means that the method is promising for multi-patch computations because the memory transfers between the main memory and the computing accelerators are often the most time-consuming part of the algorithm.

In Section \ref{sec:num_disc}, we present the numerical discretisation that we use for approximating systems of conservations. 
In Section \ref{sec:compression}, we explain how we build the compression pipeline that we use in our experiments.
In section \ref{sec:results}, we perform experiments to demonstrate the efficiency of our method.

\section{Numerical discretisation\label{sec:num_disc}}

In this section, we describe the FV and LBM schemes that we use in our experiments. The objective is to solve numerically the following system of conservation laws:
\begin{equation}
\label{eq:conservation_laws}
\partial_t W + \nabla \cdot (Q(W)) = 0.
\end{equation}
here the unknown $W(X,t)$ is a vector of $m$ conservative variables
depending of the space variable $X=(x,y)$ and the time variable $t$.
For simplicity, we assume that $X$ is in the square $]0,L[\times]0,L[$,
but more complex shapes are possible.

The divergence operator is defined by
\[
\nabla\cdot(Q(W))=\frac{\partial}{\partial x}Q^{x}(W)+\frac{\partial}{\partial y}Q^{y}(W),
\]
where $Q^{x}$ and $Q^{y}$ are two application from $\mathbb{R}^{m}$
to $\mathbb{R}^{m}$. For a given two-dimensional vector $N=(N_{x},N_{y})$,
we define the flux of the system of conservation laws by
\begin{equation}
\label{eq:flux_def}
Q(W,N)=N_{x}Q^{x}(W)+N_{y}Q^{y}(W).
\end{equation}
Finally, we define the four directions
\begin{equation}
\label{eq:directions}
N^{0}=\left(\begin{array}{c}
1\\
0
\end{array}\right),\quad N^{1}=\left(\begin{array}{c}
-1\\
0
\end{array}\right),\quad N^{2}=\left(\begin{array}{c}
0\\
1
\end{array}\right),\quad N^{3}=\left(\begin{array}{c}
0\\
-1
\end{array}\right).
\end{equation}
In this way:
\[
Q(W,N^{0})=Q^{x}(W),\quad Q(W,N^{1})=-Q^{x}(W),
\]
\[
Q(W,N^{2})=Q^{y}(W),\quad Q(W,N^{3})=-Q^{y}(W).
\]
Let $n_{X}$ be a positive integer. We approximate $W$ on a regular
grid of step 
\[
h=\frac{L}{n_{X}}.
\]
The grid is made of little square cells
\[
C_{i,j}=]ih,(i+1)h[\times]jh,(j+1)h[,\quad i=0,\ldots,n_{X}-1,\quad j=0,\ldots,n_{X}-1.
\]
For simplicity, we can assume a periodicity condition
\[
i+n_{X}\equiv i,\quad j+n_{X}\equiv j,
\]
which allows to extending the grid to any couple $(i,j)\in\mathbb{Z}\times\mathbb{Z}.$
We could also apply boundary conditions on the boundary cells. The centers of the cells are the points
\[
X_{i,j}=\left(\begin{array}{c}
x_{i}\\
y_{j}
\end{array}\right)=\left(\begin{array}{c}
ih+h/2\\
jh+h/2
\end{array}\right).
\]
The conservative data are approximated at times $t=n\tau$, at the center of the cells 
\begin{equation}
W_{i,j}^{n}\simeq W(X_{i,j},p\tau).\label{eq:approx}
\end{equation}
The system of conservation laws (\ref{eq:conservation_laws}) is then approximated by the following FV scheme, which allows computing the value at time step $n+1$ from that of the time step $n$
\begin{equation}
    \label{eq:vf_scheme}
W_{i,j}^{n+1}=W_{i,j}^{n}-\frac{\tau}{h}\sum_{k=0}^{3}Q(W_{i,j},W_{i',j'},N^{k}),\quad\text{with }\left(\begin{array}{c}
i'\\
j'
\end{array}\right)=\left(\begin{array}{c}
i\\
j
\end{array}\right)+N^{k}.
\end{equation}
In this formula, we have introduced the numerical flux $Q(W,W',N)$,
whose purpose is to approximate the flux $Q(W,N)$ at the interface
between two cells. The numerical flux has to satisfy some mathematical
property in order to ensure a stable and accurate approximation. It
is out of the scope of this work to discuss this aspect. We refer
(for instance) to \cite{eymard2000finite,leveque2002finite}.

The implementation of this algorithm is done in the following way:
\begin{itemize}
\item The data at time $n$ and $n+1$ are stored into to buffers of floats
\texttt{wn} and \texttt{wnp1}.
\item The initial data at time $t=0$ is known and allows to construct the
buffer \texttt{wn} thanks to (\ref{eq:approx}), which reads here
\[
W_{i,j}^{0}=W(X_{i,j},0).
\]
\item At each time step $n$ we compute \texttt{wnp1} from \texttt{wn} on
the full grid, thanks to (\ref{eq:fv_scheme}). These computations
make many calls to the numerical flux function $Q(W,W',N)$, which
is the heaviest part of the algorithm. It is important to take care
of the data arrangement in the buffers \texttt{wn} and \texttt{wnp1}
in order to ensure optimal memory access.
\item At the end of the time step wnp1 is copied into wn, before the next
step.
\item When the final time $T$, is reached, i.e. when $n\tau\geq T$, the
results are displayed.
\end{itemize}

\section{compression\label{sec:compression}}
In this section, we explain how we build the compression pipeline that we use in our experiments.
The compression pipeline is comparable to that of the JPEG2000 standard \cite{jpeg2000}.
The goal of this pipeline is to apply a pretreatment to the data before applying a lossless compression algorithm.
Applying a pretreatment allows us to take advantage of the spatial structure of the data to make some symbols (in this case: zeros) more frequent.

The compression pipeline is composed of three steps.
In the first step, a Discrete Wavelet Transform (DWT) is applied to the data (Subsection \ref{sec:wavelet_transform}).
The purpose of this step is to create numerous near-zeros without compromising the information content.
In the second step, a threshold is applied to the resulting DWT coefficients (Subsection \ref{sec:thresholding}).
The coefficients whose absolute value is less than the threshold are set to zero.
This step introduces the loss of information in the compression pipeline.
The loss increases as the threshold increases.
The final step is the application of a lossless compression algorithm to the thresholded DWT coefficients (Subsection \ref{sec:lossless_compression}).
This step achieves effective memory compression.
It is anticipated that the thresholding step will increase the frequency of the "zero" symbol, thus enabling the lossless compression algorithm to achieve a better compression ratio.

In the following Subsection \ref{sec:wavelet_transform}, we describe the used discrete wavelet transform.
In Subsection \ref{sec:thresholding}, we present the thresholding step.
In Subsection \ref{sec:lossless_compression}, we describe two different lossless compression algorithms that we use in our experiments.

\subsection{Wavelet Transform} \label{sec:wavelet_transform}

The Discrete Wavelet Transform (DWT) is a mathematical tool that is used to decompose a signal into a set of coefficients.
It is a discretisation of the Continuous Wavelet Transform (CWT).
The idea is to decompose a sampled signal into a sum of wavelets.
For a very general introduction to wavelet theory and applications, we refer to the book of Mallat \cite{mallat1999wavelet}.

In this work, we refer to the Battle and Lemarié (BL) wavelets \cite{dau92}

The BL wavelets, also known as the 5/3 biorthogonal wavelets, were introduced by Battle and Lemarié and are known for their ability to achieve first-order compression, which means that they filter out linear polynomials.
This makes them well suited for a wide range of applications, including image and audio compression.
They are particularly effective at preserving the important structure of a signal while achieving high compression ratios.
They are also efficient for compressing discontinuous data, because of their short filters that concentrate the compression analysis on small regions of the signal.


In this work, we use a variation of the BL5/3 wavelets for the non-periodic case and a discrete signal of length of $2^k+1$, $k \in \mathbb{N}$.
This variation is based on a folding of the wavelet (see \cite{mallat1999wavelet} p. 321).
This allows for improving the compression ratio in the non-periodic case.
In addition, we show that this construction satisfies a fundamental conservation property, which is mandatory for ensuring the convergence of the numerical scheme under grid refinement \cite{hou1994nonconservative}.

In practice, our implementation is based on the wavelet lifting approach introduced by Sweldens \cite{sweldens1995lifting, sweldens2000building}.

\subsubsection{Discrete wavelet transform in the periodic case} \label{sec:periodic_case}

Let us first introduce the wavelet compression scheme for a 1-periodic function $f$ on the real line.
Its definition on the interval $[0,1]$ is thus sufficient.

We define the following sampling points in $[0,1]$:
\[
x_{j,k}=k2^{-j},\quad0\leq k<2^{j},\quad j\geq0.
\]

The $j$ index is the scale index.
The lower scales correspond to the coarser grids and the higher scales to the finer grids.
In some sense, the scale index $j$ gives the highest frequency that can be represented by the grid $\{x_{j,k}\}_{0\leq k<2^{j}}$.
Let us also remark that the grid at scale $j$ has $2^{j}$ points.
The first point is always $0$ and the last point is always $1-2^{-j}$:
\[
x_{j,0}=0,\quad x_{j,2^{j}-1}=1-2^{-j}.
\]

For a a given scale $j=j_{0}\geq1$ the function is sampled at the grid points
\[
s_{j,k}\simeq f(x_{j,k}).
\]
The idea of the wavelet transform is to transform the set of samples $(s_{j,k})_{0\leq k<2^{j}}$ at scale $j$ into a set of samples $(s_{j-1,k})_{0\leq k<2^{j-1}}$ at the coarser scale $j-1$ and details $(d_{j,k})_{0\leq k<2^{j-1}}$.
The role of the details is to allow the reconstruction of the $s_{j,k}$'s from the $s_{j-1,k}$'s because there is a loss of information in the coarsening process.
Let us define
\[
u=\left(\begin{array}{c}
s_{j,0}\\
s_{j,1}\\
\vdots\\
s_{j,2^{j}-1}
\end{array}\right)
\]
and
\[
v=\left(\begin{array}{c}
s_{j-1,0}\\
d_{j-1,0}\\
s_{j-1,1}\\
d_{j-1,1}\\
\vdots\\
s_{j-1,2^{j-1}-1}\\
d_{j-1,2^{j-1}-1}
\end{array}\right).
\]

A step of the wavelet transform can be defined in matrix form
\begin{equation}
    v=Au.\label{eq:DWT_matrix_form}
\end{equation}

The even rows of $A$ correspond to the $s_{j-1,k}$'s while the odd rows correspond to the $d_{j-1,k}$'s.
The resulting $v$ vector gives us the grid at scale index $j-1$ as well as the details.
The details allow us to reconstruct the exact $s_{j,k}$'s from the $s_{j-1,k}$'s.
If the DWT scheme is properly designed, the transformation is bijective and there is an $A^{-1}$ matrix such that

\[
u=A^{-1}v.
\]

\subsubsection{Wavelet transform in the non-periodic case}

We now adapt the previous scheme to the non-periodic case.
The difference with the periodic case is that we have to take into account the boundaries of the interval.
We propose a wavelet construction where the wavelets filter out linear polynomials.

We consider a function on the interval $[0,1]$.
We define the following sampling points
\[
x_{j,k}=k2^{-j},\quad0\leq k\leq2^{j},\quad j\geq0.
\]

At scale $j$, the signal is represented by the grid  $\{x_{j,k}\}_{0\leq k\leq2^{j}}$ and has $2^{j}+1$ points.
The first point is always $0$ and the last point is always $1$:
\[
x_{j,0}=0,\quad x_{j,2^{j}}=1.
\]
They both correspond to even index points.
The usual wavelet constructions are on the whole real line or on a periodic
domain.
The separation between even or odd indices $k$ is very important.
In the usual wavelet constructions, there are as many odd points as even points.
Because we are on an interval, at scale $j\geq1$ we have $2^{j-1}+1$ even indices and $2^{j-1}$ odd indices.

For a given scale $j=j_{0}\geq1$ the function is sampled at the grid points
\[
s_{j,k}\simeq f(x_{j,k}).
\]

At coarser scales, we decide to always keep the boundary values unmodified
\begin{equation}
s_{j-1,0}=s_{j,0}=f(0),\quad s_{j-1,2^{j-1}}=s_{j,2^{j}}=f(1).\label{eq:boundary_values}
\end{equation}

The idea of having different resolutions through the $j$ scales is called the multiresolution analysis.
This technique is aimed to represent the samples more efficiently by keeping only the relevant information at scale $j<j_{0}$.
Another aspect of the multiresolution analysis is that it gives a way to extrapolate $f$ at finer scales $j>j_{0}$ on the dyadic points $x_{j,k}$.

If the signal represented by $f$ has smooth variations, it seems reasonable to keep only the even samples $s_{j,2k},\quad0\leq k\leq2^{j-1}$.

By linear interpolation, we expect that the odd samples satisfy
\[
s_{j,2k+1}\simeq\frac{s_{j,2k}+s_{j,2(k+1)}}{2},\quad0\leq k\leq2^{j-1}-1.
\]
Because it is not exact, we keep track of the details at the scale $j-1$
\[
d_{j-1,k}=s_{j,2k+1}-\frac{s_{j,2k}+s_{j,2(k+1)}}{2},\quad0\leq k\leq2^{j-1}-1,
\]
which allows us to exactly reconstruct the initial odd values at the scale $j$:
\[
s_{j,2k+1}=d_{j-1,k}+\frac{s_{j,2k}+s_{j,2(k+1)}}{2},\quad0\leq k\leq2^{j-1}-1.
\]

Now, we introduce a mass conservation requirement:
\begin{equation}
    \frac{f(0)+f(1)}{2}+\sum_{k=1}^{2^{j-1}-1}s_{j-1,k}=\frac{f(0)+f(1)}{4}+\frac{1}{2}\sum_{k=1}^{2^{j}-1}s_{j,k}.\label{eq:conservation}
\end{equation}
Essentially, this is a quadrature formula that states that the mass of the samples at the scale $j$ equals the mass of the samples at the scale $j-1$.
We use the trapezoidal quadrature formula.
This explains why the weights of the first and last samples are halved.
The mass conservation property (\ref{eq:conservation}) is an essential property for ensuring the convergence of the whole numerical scheme.
It ensures that the mass of the signal $s_{j,k}$ is concentrated in the samples $s_{j-1,k}$.
This implies that the details $d_{j-1,k}$ wear no mass.
Therefore, the detail coefficients $d_{j-1,k}$ can be modified without affecting the mass of the signal.

The last equation that needs to be set is the relation between $s_{j-1,k}$ and $s_{j,2k}$.
The naive relation $s_{j-1,k}=s_{j,2k}$ does not always verify the mass conservation property (\ref{eq:conservation}).
To satisfy this property, we introduce $\alpha_{j,k}$ coefficients and we set
\[
s_{j-1,k}=s_{j,2k}+\alpha_{j-1,k-1}d_{j-1,k-1}+\alpha_{j-1,k}d_{j-1,k},\quad0<k<2^{j-1}.
\]
This equation corresponds to the lifting part of the wavelet construction introduced by Sweldens \cite{sweldens1995lifting, sweldens2000building}.
This formula has no meaning if $k=0$ or $k=2^{j-1}$, but then we use (\ref{eq:boundary_values}) to find that $s_{j-1,0}=s_{j,0}$ and $s_{j-1,2^{j-1}}=s_{j,2^{j}}$.
With the constraints we set, one can verify that the only choice of coefficients that verifies the mass conservation property (\ref{eq:conservation}) is
\begin{equation}
    \alpha_{j-1,k}=
        \begin{cases}
            \frac{1}{4}    & \text{if } 0<k<2^{j-1}-1\\
            \frac{1}{2}    & \text{if } k=0\text{ or } k=2^{j-1}-1
        \end{cases}
\end{equation}

The presented formulae let us perform one DWT step.
Multiple steps can be performed by applying the same scheme to the scales $j-1$, $j-2$, $\ldots$
At the end of the process, we have $2^{j-L}+1$ samples and $2^{j}-2^{j-L}$ details, where L is the number of performed DWTs, also known as the compression level.
If most details are near-zeros, we can expect the compression ratio to increase as the compression level $L$ increases.

This scheme can be expressed in a matrix form with (\ref{eq:DWT_matrix_form}).
Because of our construction, the $v$ vector has $2^{j-1}+1$ samples and $2^{j-1}$ details while the one presented in \ref{sec:periodic_case} has $2^{j-1}$ samples.
For $j=3$, we obtain
\[
A=\frac{1}{8}\left[\begin{array}{ccccccccc}
8 & 0 & 0 & 0 & 0 & 0 & 0 & 0 & 0\\
\noalign{\medskip}-4 & 8 & -4 & 0 & 0 & 0 & 0 & 0 & 0\\
\noalign{\medskip}-2 & 4 & 5 & 2 & -1 & 0 & 0 & 0 & 0\\
\noalign{\medskip}0 & 0 & -4 & 8 & -4 & 0 & 0 & 0 & 0\\
\noalign{\medskip}0 & 0 & -1 & 2 & 6 & 2 & -1 & 0 & 0\\
\noalign{\medskip}0 & 0 & 0 & 0 & -4 & 8 & -4 & 0 & 0\\
\noalign{\medskip}0 & 0 & 0 & 0 & -1 & 2 & 5 & 4 & -2\\
\noalign{\medskip}0 & 0 & 0 & 0 & 0 & 0 & -4 & 8 & -4\\
\noalign{\medskip}0 & 0 & 0 & 0 & 0 & 0 & 0 & 0 & 8
\end{array}\right]
\]
and
\[
A^{-1}=\frac{1}{8}\left[\begin{array}{ccccccccc}
8 & 0 & 0 & 0 & 0 & 0 & 0 & 0 & 0\\
\noalign{\medskip}4 & 6 & 4 & -1 & 0 & 0 & 0 & 0 & 0\\
\noalign{\medskip}0 & -4 & 8 & -2 & 0 & 0 & 0 & 0 & 0\\
\noalign{\medskip}0 & -2 & 4 & 6 & 4 & -1 & 0 & 0 & 0\\
\noalign{\medskip}0 & 0 & 0 & -2 & 8 & -2 & 0 & 0 & 0\\
\noalign{\medskip}0 & 0 & 0 & -1 & 4 & 6 & 4 & -2 & 0\\
\noalign{\medskip}0 & 0 & 0 & 0 & 0 & -2 & 8 & -4 & 0\\
\noalign{\medskip}0 & 0 & 0 & 0 & 0 & -1 & 4 & 6 & 4\\
\noalign{\medskip}0 & 0 & 0 & 0 & 0 & 0 & 0 & 0 & 8
\end{array}\right].
\]

In the even rows of $A$, we can read the wavelet low-pass filter coefficients.
In the odd rows, we read the high-pass filters.
Only the filters at the beginning and at the end of the interval have small variations.
In the middle they are constant.
Away from the boundaries, we recognise the filters of the first order 5/3 biorthogonal wavelet transform presented in \cite{cohen1992biorthogonal}.
Those wavelets are attributed to Battle and Lemarié \cite{dau92} and are usually referred to as the BL5/3 wavelets.

The presented wavelet transform achieves first order compression, meaning that it filters out linear polynomials.
It is possible to achieve order two compression by using the Cohen-Daubechies-Feauveau 9/7 wavelets \cite{cohen1992biorthogonal}.
However, increasing the order of the filtered polynomials is not necessarily significantly better in terms of compression ratio.
In this work, we will only use the BL5/3 wavelets.

\begin{figure}[htbp]
    \centering
    \begin{subfigure}[b]{0.35\textwidth}
        \includegraphics[width=\textwidth]{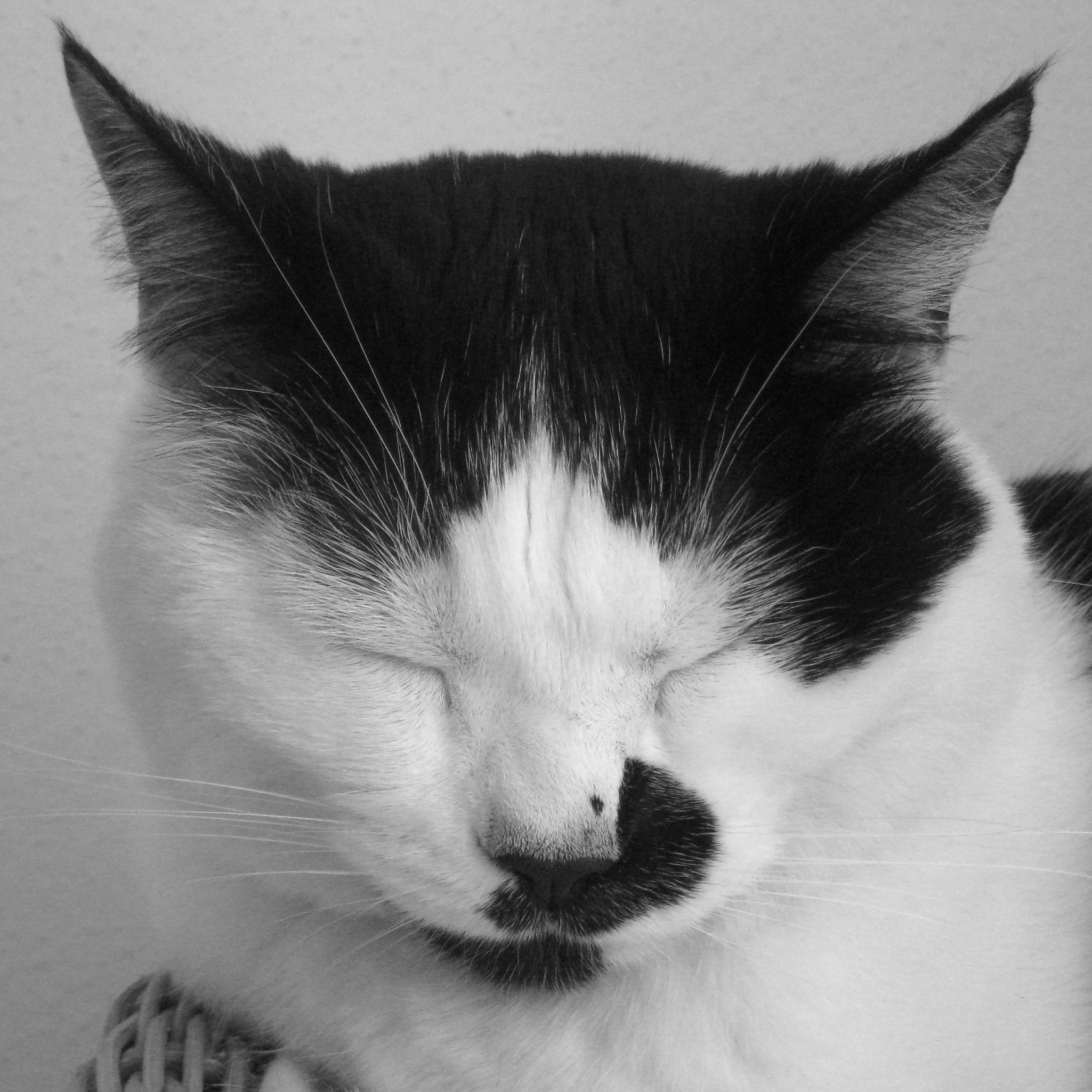}
        \caption{Original grayscale image.}
        \label{fig:miaou}
    \end{subfigure}
    \begin{subfigure}[b]{0.35\textwidth}
        \includegraphics[width=\textwidth]{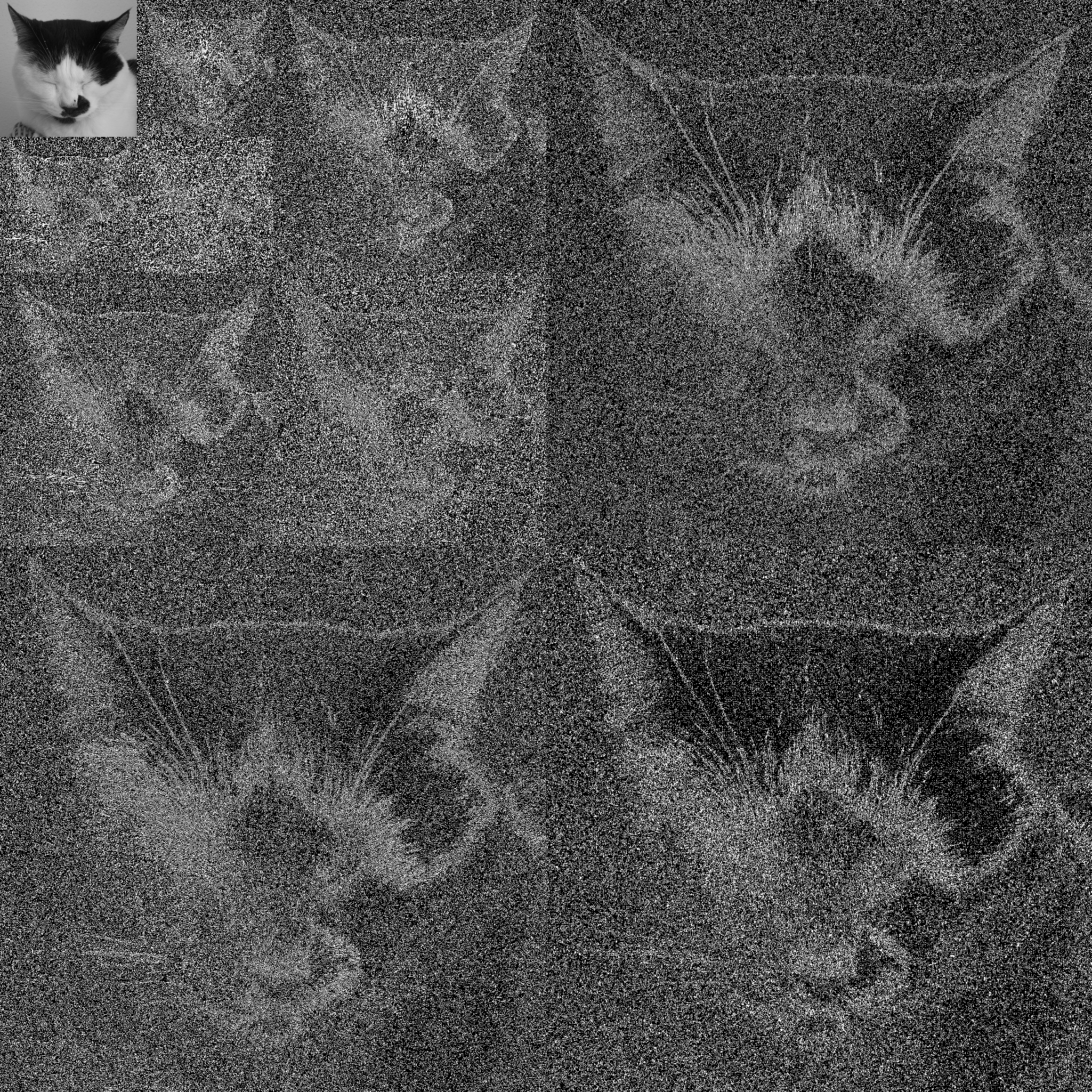}
        \caption{Resulting grayscale image.}
        \label{fig:miaou-3step}
    \end{subfigure}
    \caption{Transformation of a 2-dimensional image (left) after the application of 3 steps of the wavelet scheme (right). In the result (right), the samples are located in the upper-right corner of the image. The remaining coefficients represent details at different levels. Details colors are obtained with $\text{mod}(x,256)$, where $x$ is the detail value. The resulting color ranges from 0 (black) to 255 (white).}
    \label{fig:miaou-2d}
\end{figure}

This section presents a 1-dimensional wavelet transform.
To extend it to N dimensions, we simply apply this scheme successively to each dimension.
For example, in 2 dimensions (a matrix), we first apply the wavelet transform to all the rows and then to all the columns.
Figure \ref{fig:miaou-3step} shows the result of the application of 3 steps of the wavelet scheme to a 2-dimensional image (Figure \ref{fig:miaou}).
We can recognise the original image in the samples.
We can also recognise a sketchy version of the image in the different detail levels.
In particular, we can see that the outline of the cat is well distinguishable in the details.
We can see that the sought property is reached: the parts of the image that are near-linear result in small coefficients (very dark or very bright colors), while the non-linear parts (outline) result in large coefficients.

\begin{figure}[htbp]
    \centering
    \begin{subfigure}[b]{0.49\textwidth}
        \includegraphics[width=\textwidth]{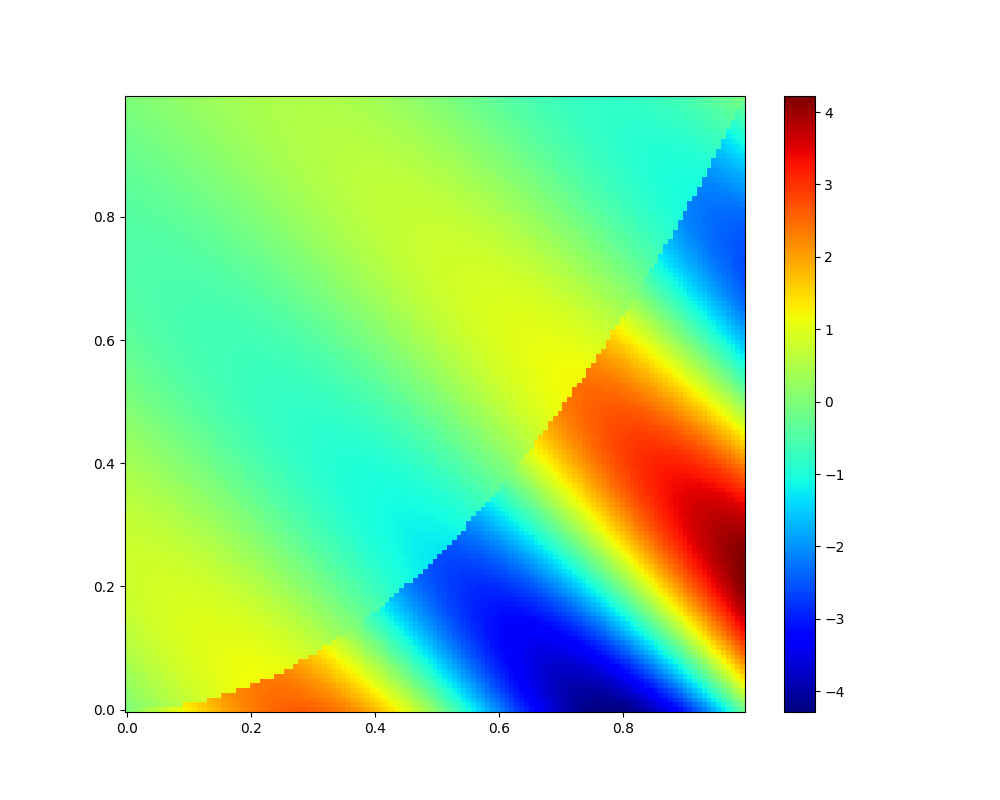}
        \caption{Original function.}
        \label{fig:regular}
    \end{subfigure}
    \begin{subfigure}[b]{0.49\textwidth}
        \includegraphics[width=\textwidth]{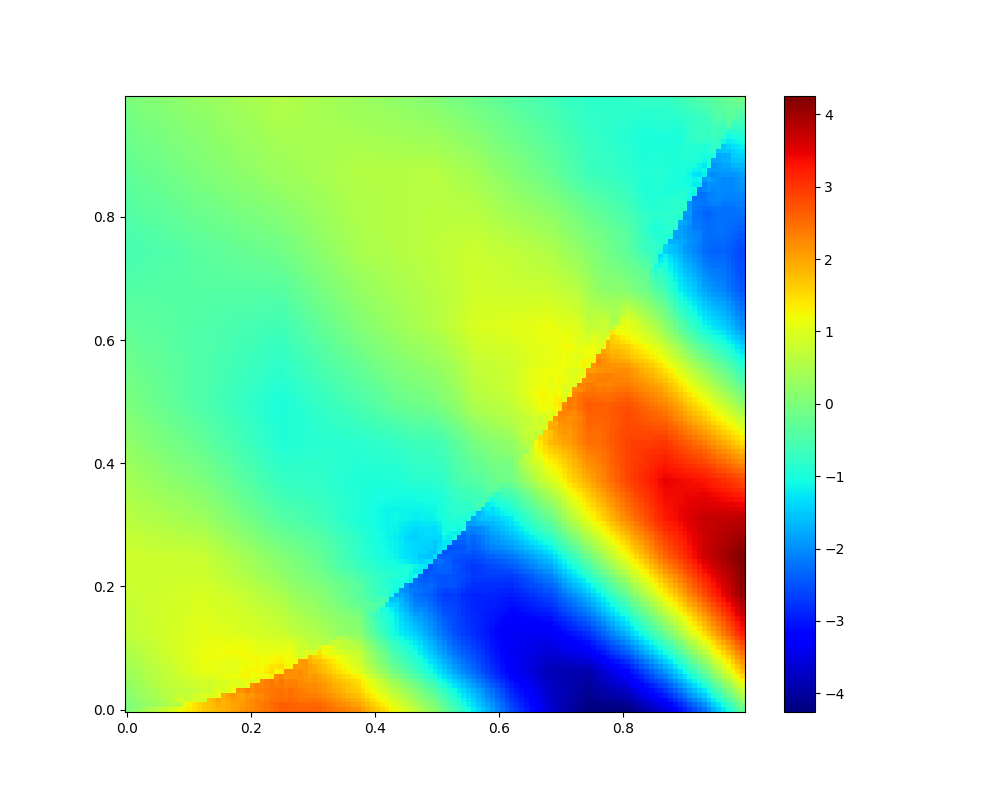}
        \caption{Reconstructed function.}
        \label{fig:regular-6step}
    \end{subfigure}
    \caption{Transformation of a 2-dimensional regular function with a discontinuity on a 129x129 grid (left) after the application of 6 steps of the wavelet scheme (right).
    The details that are between $-0.2$ and $0.2$ are nullified to create a loss.
    The reconstructed image (right) is noticeably degraded.
    In particular, artefacts appear near the discontinuity.
    }
    \label{fig:regular-2d}
\end{figure}

To demonstrate the ability of the scheme to preserve discontinuities, we present the result of the application of 6 steps of the wavelet scheme to a 2-dimensional regular function that has a discontinuity.
The function is defined by:
\[
    f(x,y)=e^{x-y}\sin(2\pi(x+y))\times \text{step}(y-x^2),
\]
where
\[
    \text{step}(x)=\begin{cases}
        1 & \text{if } x \geq 0\\
        2 & \text{if } x < 0
    \end{cases}.
\]
Figure \ref{fig:regular-2d} shows the plots before and after the application of the wavelet scheme.
All the details that are between $-0.2$ and $0.2$ are nullified to create a loss.
The reconstructed image (Figure \ref{fig:regular-6step}) is close to the original function.
In particular, the discontinuity is well preserved.
This is due to the fact that large details (induced by the discontinuity) are fully kept and let us perfectly reconstruct the original signal.
We can also notice that artefacts appear near the discontinuity.
These artefacts are typically due to the thresholding of a low-level detail.
We verify experimentally that the mass conservation property is reached.
The masses of the signal before and after the transformation are equal (up to the machine precision level), even when all the details are nullified.
Incidentally, the number of non-null coefficients in the compressed form is 481 (out of 16641 in total), meaning that we can reasonably expect a compression ratio of the order of $\frac{481}{16641}\approx 34.596674$.

\subsection{Thresholding} \label{sec:thresholding}

In the thresholding step, the goal is to transform the near-zeros values into zeros depending on a threshold.
This threshold depends on the scale $j$ and is only applied to the details.
In the case of an N-dimensional wavelet transform, the threshold depends on the N-tuple $(j_1,\ldots,j_N)$ formed by the scales of each dimension.
We use a constant $c$ to have a global control on the threshold.
This constant represents the threshold value for the lowest scale (coarsest) details.
We define 3 thresholding functions:
\begin{itemize}
\item \textbf{Constant thresholding}: the threshold is constant along all the scales. $(j_1,\ldots,j_N)\mapsto c$.
\item \textbf{Accumulation thresholding}: the threshold is defined by $(j_1,\ldots,j_N)\mapsto c\times\alpha^{\sum -j_i}$ where $\alpha$ is a constant that we set to 2 and $j_i$ are the scales for the N dimensions (in 2 dimensions, $j_x$ and $j_y$ for example). With this method, the coarser scales have a higher threshold and are, therefore, less likely to be nullified.
\item \textbf{Capped thresholding}: the threshold is defined by $(j_1,\ldots,j_N)\mapsto c\times\alpha^{max(j_i)}$. Again, we set $\alpha$ to 2. This method is similar to the accumulation thresholding, but it only takes into account the highest scale between each dimension.
\end{itemize}
In essence, these different methods aim at setting a fair threshold.
The constant thresholding is the simplest one and assumes that nullifying an equally-valued detail on two different scales has the same impact.
The accumulation thresholding and the capped thresholding try to take into account the fact that the coarser scales may have a larger impact on the final result.
These different methods correspond to the type of loss that we want minimise.
To justify the use of one method over the other, we test them empirically.
For this, we set an arbitrary maximum accepted error for a simulation and manually tune $c$ to achieve this error.
Then, the most efficient method is the one that achieves the best compression ratio, i.e. the one that nullifies the most details (for the same error).
According to this test, the capped thresholding is the most efficient and will be the one used in the following experiments.
The $c$ constant will be referred to as the \textit{threshold value} and can be viewed as the used threshold for the first level of details.

After the thresholding step, the small details are set to zero but still stand in the memory.
Thus, the data size is not changed.
An additional lossless compression step is necessary to effectively compress the data.

\subsection{Lossless compression} \label{sec:lossless_compression}

In the context of GPU computations, the use of lossless data compression techniques is often explored.
These are particularly relevant when the data need to be transferred between the CPU and the GPU.

For instance, Patel \textsl{et al.} propose a fast compression algorithm for GPU computations \cite{parallel_lossless_compression}.
Their algorithm is based on the bzip2 compression algorithm.
E. Sitaridi \textsl{et al.} develop a fast GPU decompression method based on the LZ4 method \cite{parallel_lossless_decompression, LZ4_yann_collet}.
F. Knorr \textsl{et al.} propose a fast lossless GPU compression method for floating-point scientific data known as ndzip-gpu \cite{knorr2021ndzip}.
This compression method typically achieves excellent compression throughput but worse compression rate.

Currently, many GPU compression algorithms are based on the use of the nvCOMP library \cite{nvcomp_2020}.
This library has become widely used in the HPC community and provides a lossless compression/decompression framework that aims at being runtime-efficient.
It is based on the use of the CUDA API and is compatible with the CUDA programming model.
Multiple highly efficient compression algorithms are implemented in this library, such as LZ4, zStandard, or Bitcomp.

In the case of FV/LBM simulations, the use of these generic lossless compressions would typically not be the most adapted as they would not take into account the spatial structure of the data.
Moreover, it can be acceptable to use a lossy compression method, as long as we can verify that the induced loss does not impact the validity of the simulation.
This is why we combine a Wavelet Lifting Scheme with a lossless compression algorithm.

In this work, we compare two different lossless compression’s schemes: a CSR (Compressed Sparse Row) method provided by cuSPARSE and an LZ4 method provided by the nvCOMP library.

These lossless compressions are performed after the previous wavelet transform step to achieve an effective compression of the data.

\subsubsection{Sparse Matrix Representation}

The Compressed Sparse Row (CSR) format, also known as the Yale format, is a widely used representation for sparse matrices.
In this format, the non-zero elements of the matrix are stored in 3 arrays.
The first array, $V$, contains the non-zero elements of a $m\times n$ matrix.
The second array, $COL$, of the same size as $V$, contains the column indices of the non-zero elements.
The third array, $ROW$, of size $m+1$ contains the offset of the first non-zero element of each row.
This representation can be used as a lossless compression technique since we anticipate a large number of zero values in the matrix.
While the CSR format was not designed specifically for compression purposes, it is both well-known and efficient.
Additionally, its compressed size is proportional to the number of non-zero elements, making it a consistent representation.

\subsubsection{LZ4}

LZ4 is a widely used lossless compression algorithm known for its high compression ratios and fast processing speeds on GPU.
It has gained popularity due to its efficient CUDA implementation in the nvCOMP library, which was developed by Nvidia \cite{nvcomp_2020}.
LZ4 is a byte-oriented algorithm that is designed to be fast and parallelizable.
It was initially developed to perform well on CPUs \cite{LZ4_CPU} and has since been optimised for use on GPUs.
The algorithm uses a block-based approach and compresses each block independently, with a configurable chunk size.
The chunk size determines the size of the input data that is processed at once by the LZ4 algorithm.
Usually, a larger chunk size implies a better compression ratio, but a slower compression speed.
LZ4 is commonly used in data-intensive applications such as scientific simulations and big data analytics.

\section{Practical GPU implemention} \label{sec:results}

\subsection{Protocol}

We propose a numerical experiment to test the performance of our algorithm.
The idea is to perform a two-dimensional LBM/FV simulation while introducing compression/decompression cycles between each time step.

\subsubsection{Structure of the data}

\begin{figure}[h]
	\centering
	\includegraphics[scale=0.4]{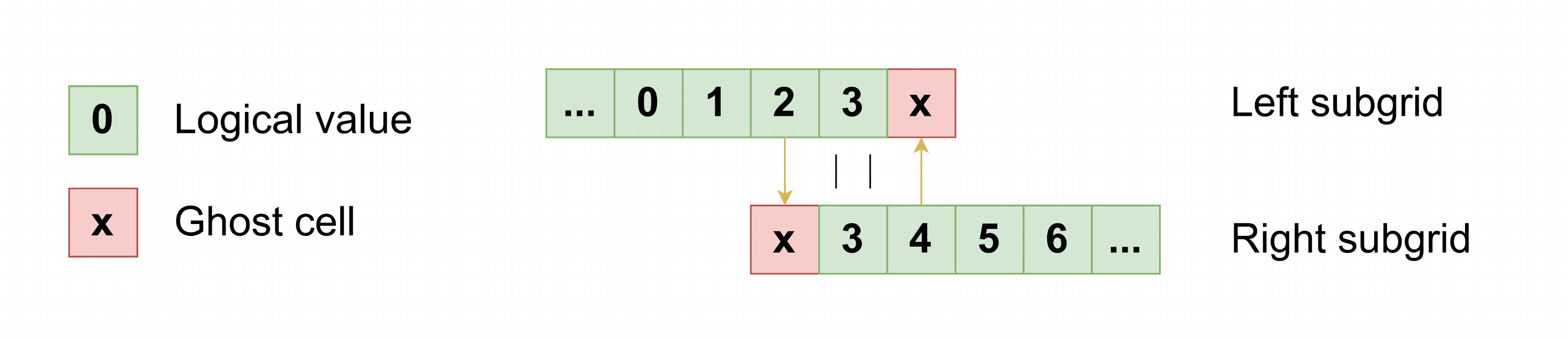}
	\caption{
	    This schema shows the synchronisation process in 1 dimension.
        The ghost cells are updated by copying the corresponding values of the neighbouring subgrids.
        One value is shared between both subgrids and does not need to be updated.
        The yellow arrows show which values are copied.
        In a real-case scenario, the data consist of floating-point values, rather than integers.
        In multiple dimensions, the synchronisation process is performed successively in each dimension.
	}
	\label{fig:synchronization_process}
\end{figure}

The simulation grid is divided into 8 subgrids: 2 subgrids in each of the 3 dimensions.
This division in subgrids does not serve any computational purpose but forces the wavelet transform to be applied on a non-periodic space.
Each subgrid has a logical size and a true size.
The logical space represents the simulation space, meaning that each value corresponds to a fixed space in the simulation.
The true size is the whole space of the subgrid. It contains the logical space and the ghost cells (also known as halo or overlap).
The ghost cells are useful in a simulation because they duplicate the values of the neighbouring cells and avoid the need to explicitly communicate with the neighbouring subgrids.
Keeping the ghost cells up to date, however, requires a synchronisation phase between time steps.
The synchronisation process is illustrated in Figure \ref{fig:synchronization_process}.

We chose a simulation grid of size 128x128x128.
Each subgrid has a logical size of 65x65x65 and a true size of 67x67x67.
The bordering values of the logical space are shared between multiple subgrids.
In Figure \ref{fig:synchronization_process}, the shared value is the cell "3".
As these values are not synchronised, the same FV/LBM computation must be performed once for each subgrid.
The reason for introducing this overlap is to keep the mass-conservation property of the wavelet transform.
Indeed, one can verify that the mass conservation property equation (\ref{eq:conservation}) implies that:
\begin{equation}
\sum (s_{j,k}-T(s_{j,k})) \times
    \begin{cases}
        \frac{1}{2}    & \text{if } k=0\text{ or }k=2^j-1\\
        1              & \text{otherwise}
    \end{cases}
 = 0,
\end{equation}
where $s_{j,k}$ is the $k$-th coefficient, $j$ is the scale and $T$ is the application of the compression pipeline: wavelet transform, thresholding, and inverse wavelet transform.
This means that the total mass of the compressed interval is conserved regardless of the amount of information lost during the thresholding phase.
By sharing a bordering value, we ensure that this exact mass conservation property is achieved for the entire grid.

\subsubsection{Transport simulation}

\begin{figure}[h]
    \centering
    \begin{subfigure}[b]{0.35\textwidth}
        \centering
        \includegraphics[width=\textwidth]{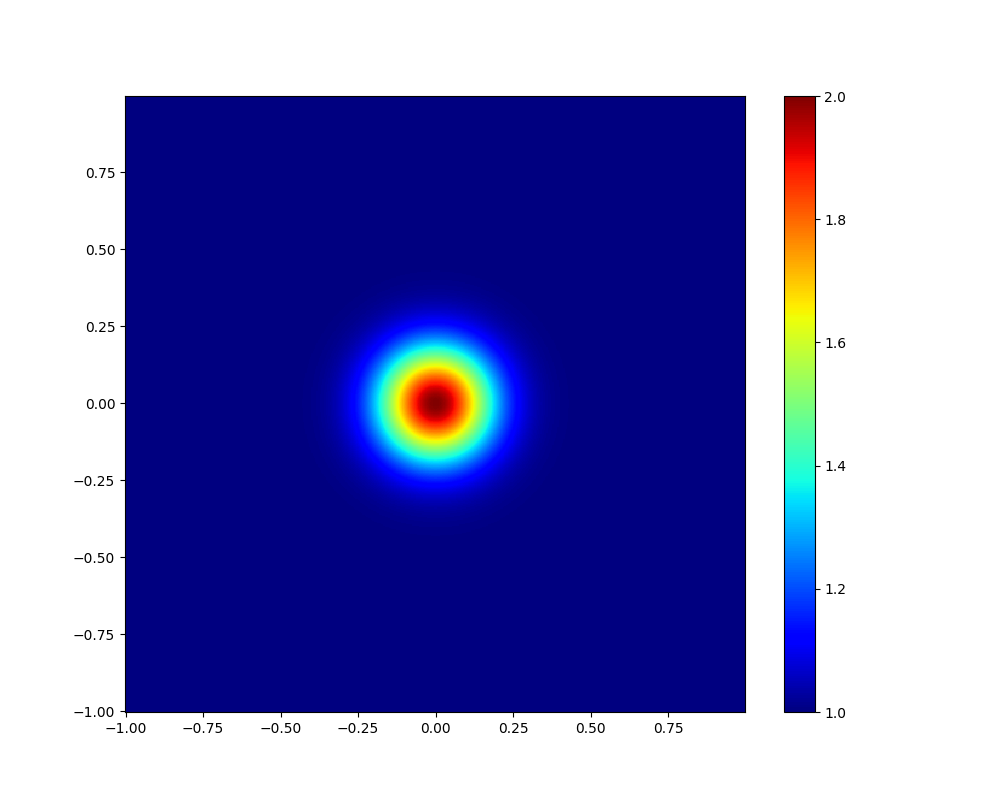}
        \caption{t=0s}
        \label{fig:LBM_simulation_1}
    \end{subfigure}
    \begin{subfigure}[b]{0.35\textwidth}
        \centering
        \includegraphics[width=\textwidth]{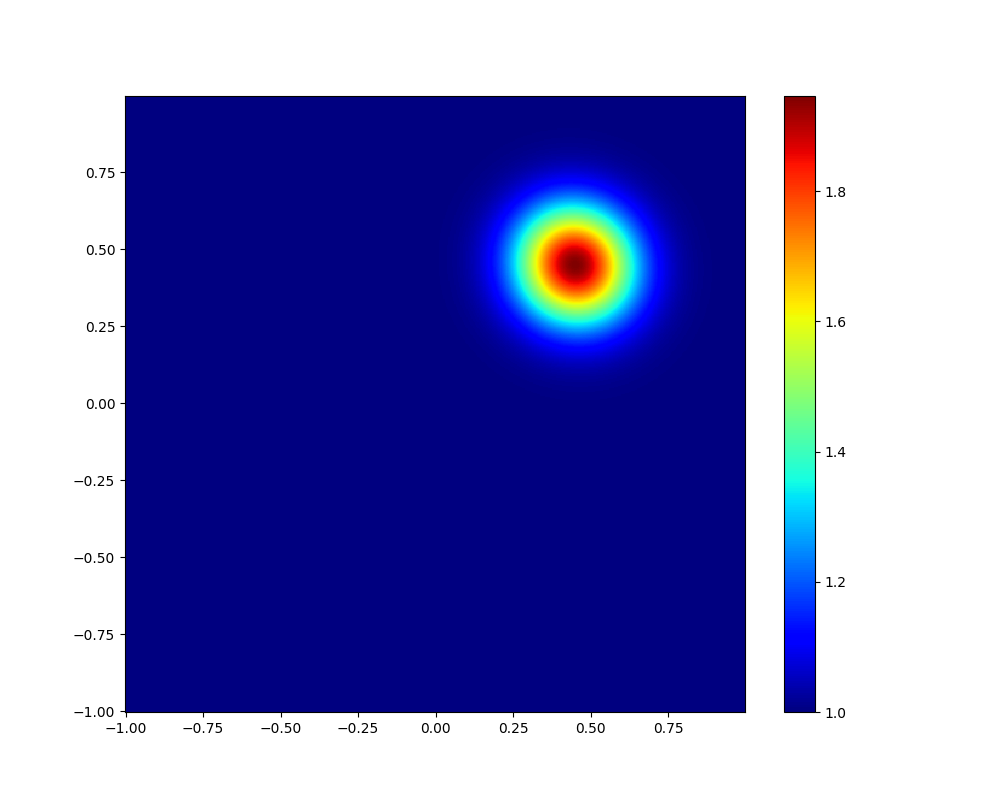}
        \caption{t=0.5s}
        \label{fig:LBM_simulation_2}
    \end{subfigure}
    \caption{
        This figure shows the initial state of the simulation (left) and the exact solution of the simulation at t=0.5s for $\alpha=0.9$ and $\beta=0.9$ (right).
        The original structure is displaced with a speed of ($\alpha$, $\beta$).
        The color represents the value of the function.
    }
    \label{fig:LBM_simulation}
\end{figure}

The first tested  simulation is a 2D simplistic computation of the displacement of an arbitrary structure given by the following rules:
\begin{equation}
\begin{cases}
f\_init(x,y) = 1 + e^{-30(x^2+y^2)}\\
f(x,y,t) = f\_init(x-\alpha t, y-\beta t).
\end{cases}
\end{equation}
In other words, $W=f\in \mathbb{R}^m$, $m=1$, is solution of the transport equation
$$
\partial_{t}W+\nabla\cdot\left(W\left(\begin{array}{c}
    \alpha\\
    \beta
    \end{array}\right)\right)=0.
$$
With the definition (\ref{eq:flux_def}), this amounts to considering the conservation law with the flux
$$
Q(W,N)=W(\alpha N^{x}+\beta N^{y}).
$$
The $f\_init$ is the initial state and $f$ is the state at time $t$.
The initial state is a 2D Gaussian function centered in the middle of the grid (Figure \ref{fig:LBM_simulation_1}).
This structure is displaced at a constant speed $(\alpha, \beta)$.
The transport equation  is solved using the scheme (\ref{eq:vf_scheme}) with the standard upwind flux

\begin{equation}
    \label{eq:upwind_trans}
    Q(W_{L},W_{R},N)=W_{L}\max(\alpha N^{x}+\beta N^{y},0)+W_{R}\min(\alpha N^{x}+\beta N^{y},0).
\end{equation}

where $\Delta x$ is the size of the cells, $\text{CFL}$ is the Courant-Friedrichs-Lewy number, $v_{max} = max(\alpha, \beta)$ is the maximum speed of the structure and $dt = \frac{\text{CFL} \cdot \Delta x}{v_{max}}$ is the duration of a time step.
This scheme can be implemented as follows:
\lstset{
  language={C++},
}

\begin{minipage}{\linewidth}
\begin{lstlisting}[caption={Pseudo-code of the transport kernel}, captionpos=b, label={lst:LBM_algorithm}, mathescape]
for i = 1 to NX-2 do
    for j = 1 to NY-2 do
        W_next[i][j] = W_now[i][j]
        for dir = 0 to 3 do
            W_next[i][j] -= dt/h * fluxnum(W_now[i][j],
                                           W_now[i+N[dir][0]][j+N[dir][1]],
                                           N[dir])
        end for
    end for
end for
\end{lstlisting}
\end{minipage}
Where \texttt{fluxnum} is a function implementing the numerical flux and \texttt{N} is the array of the normal vectors of the cells defined in (\ref{eq:directions}).

Note that i=0, j=0, i=NX-1, and j=NY-1 can be omitted as they are part of the ghost cells and will be overwritten by the neighbouring subgrids during the synchronisation phase.
It is a stencil computation, meaning that the next state of each cell (excluding the borders) is a function $\varphi$ of its neighbours.
Here, the stencil targets 5 cells: the current cell and its 4 neighbours.

Figure \ref{fig:LBM_simulation_1} shows f\_init at t=0s.
Figure \ref{fig:LBM_simulation_2} shows the exact solution at t=0.5s for $\alpha=0.9$ and $\beta=0.9$.

\subsubsection{Compression}

To test the compression algorithm, we perform a compression/decompression cycle between each time step.
The compression cycle is described in Section \ref{sec:compression}.
We recall that the compression consists of (1) applying the wavelet transforms on the subgrids, (2) thresholding near-zeros, and (3) applying a lossless compression on the resulting data.
The decompression process consists of performing the opposite operations in reverse order.
A various number of successive wavelet transforms (1) can be applied, the limit being that the number of samples $2^j+1$ must remain valid.
We exclude the ghost cells from the wavelet transform because they are not part of the logical space, meaning that the information they contain can be rediscovered thanks to the neighbouring subgrids.
The number of performed wavelet transforms is a parameter of the algorithm that we will refer to as the \textit{compression level}.
The threshold value (2) is manually set and corresponds to the $c$ constant introduced in Subsection \ref{sec:thresholding}
A threshold of 0 implies that no data is nullified.

We test two lossless compressions (3): the \textit{CSR} matrix format and the \textit{LZ4} compression.
The CSR conversion is performed using the \textit{cuSPARSE} library.
The LZ4 compression is performed using the \textit{nvCOMP} library.

\subsubsection{Methodology}

The benchmark program has been written in CUDA and compiled with the nvcc compiler and the \texttt{-O3 -use\_fast\_math} flags.
We run the program on an NVIDIA Tesla V100 GPU with 12GB of memory.

We refer to the following program parameters:
\begin{itemize}
    \item the number of performed wavelet transforms (i.e. the compression level);
    \item the threshold from which the details are nullified;
    \item the used lossless compression (LZ4 or CSR);
    \item in the case of LZ4, the chosen chunk size. A lower chunk size leads to a lower compression rate, but a higher compression speed;
    \item the measured simulation time (related to the number of time steps and the grid size).
\end{itemize}

We set two measures of interest: the effective compression ratio and the quality of the simulation.
The effective compression ratio is the data size after the compression divided by the initial data size.
The quality of the simulation is measured by comparing the results of the simulation with the exact solution.
It is measured with the L2 error against the exact solution at a given time step with the formula:
\begin{equation}
    \label{eq:L2_error}
    L2\_error = \frac{SizeX \times SizeY \times SizeZ}{NX \times NY \times NZ} \times\sum_{i=0}^{NX-1} \sum_{j=0}^{NY-1} \sum_{k=0}^{NZ-1} (f\_sim(i,j,k) - f\_exact(i,j,k))^2
    .
\end{equation}

The primary objective of this study is to determine the effect of different parameters on the compression ratio and the quality of the simulation, which is discussed in detail in Sections \ref{sec:compression_ratio_simulation}, \ref{sec:compression_ratio_threshold}, and \ref{sec:quality_simulation}.
In Section \ref{sec:computational_cost}, the computational cost of the compression/decompression cycles is analysed, serving as a preliminary reference for the performance of our implementation.

\subsection{Results}

\subsubsection{Compression ratio during the simulation} \label{sec:compression_ratio_simulation}

\begin{figure}[htbp]
    \centering
    \begin{subfigure}[b]{0.38\textwidth}
        \centering
        \includegraphics[width=\textwidth]{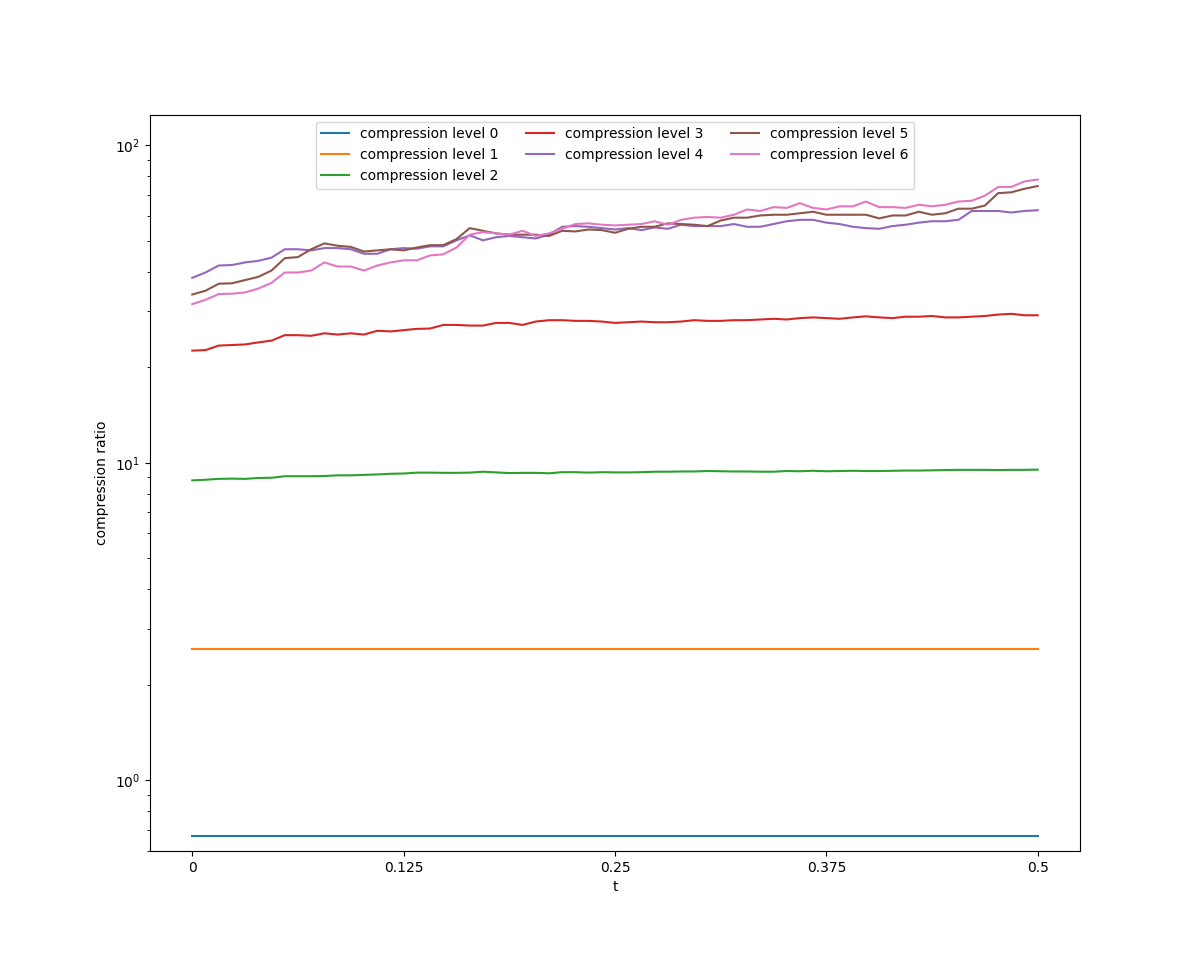}
        \caption{CSR}
        \label{fig:compression_over_time_1}
    \end{subfigure}
    \begin{subfigure}[b]{0.38\textwidth}
        \centering
        \includegraphics[width=\textwidth]{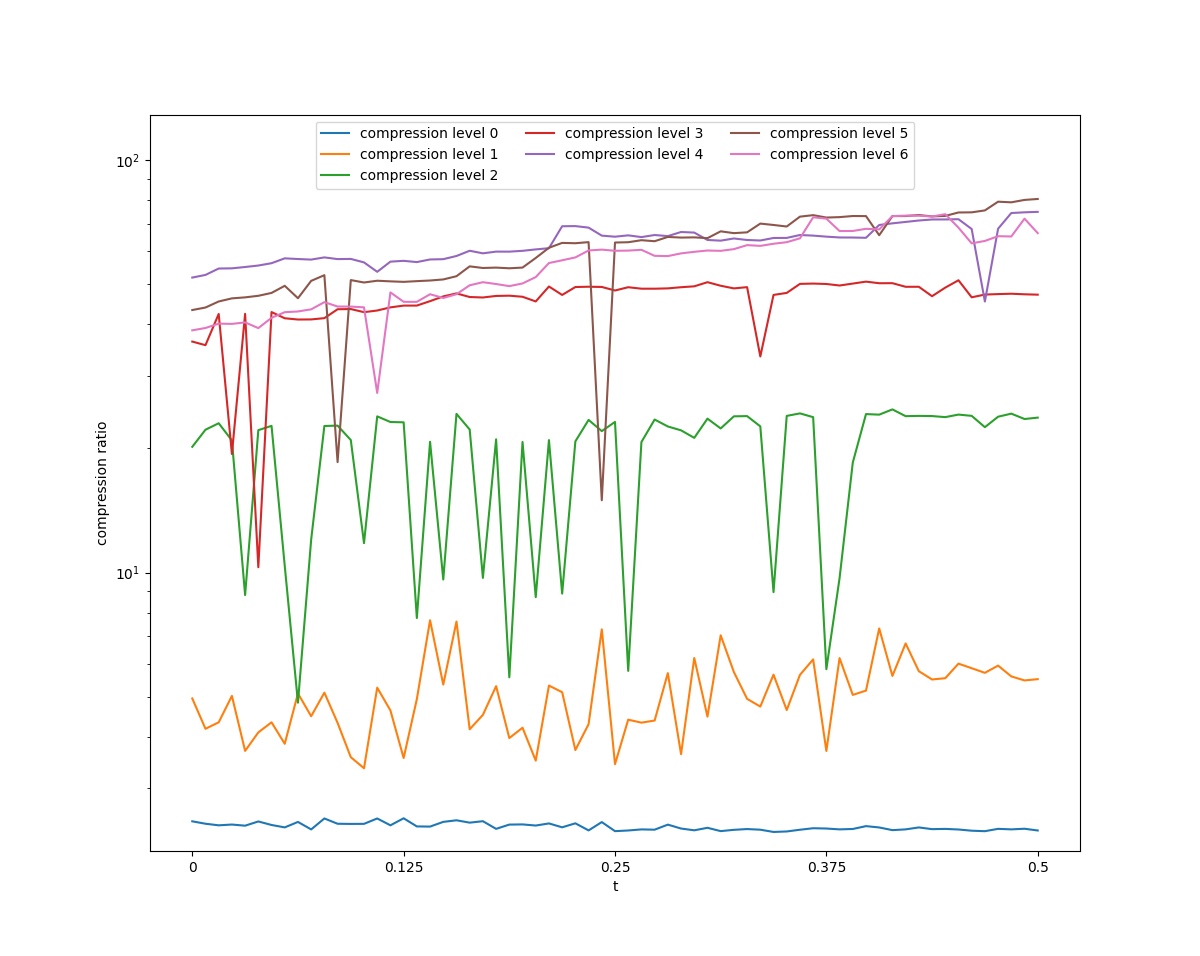}
        \caption{LZ4, chunk = 64KB}
        \label{fig:compression_over_time_2}
    \end{subfigure}
    \begin{subfigure}[b]{0.38\textwidth}
        \centering
        \includegraphics[width=\textwidth]{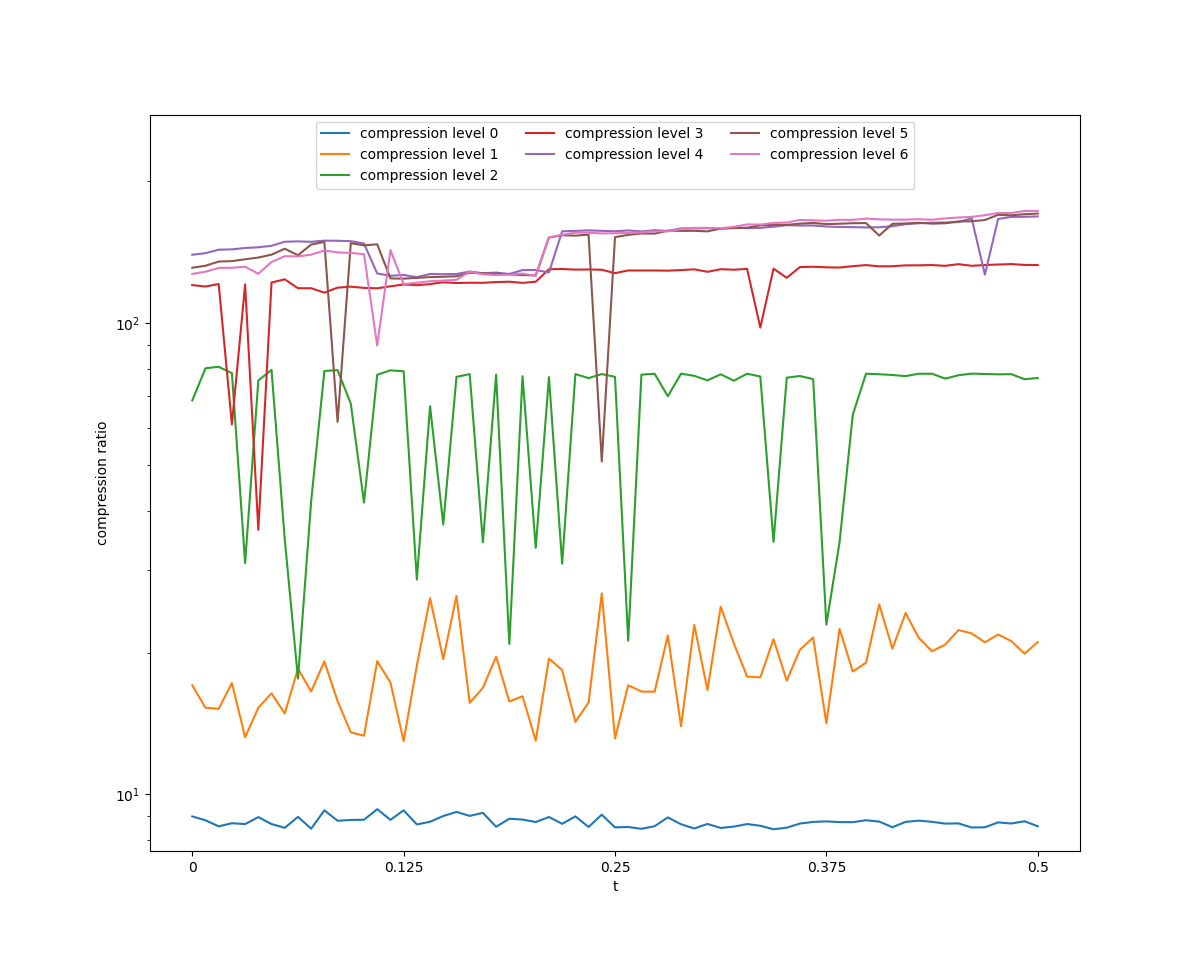}
        \caption{LZ4, chunk = 256KB}
        \label{fig:compression_over_time_3}
    \end{subfigure}
    \begin{subfigure}[b]{0.38\textwidth}
        \centering
        \includegraphics[width=\textwidth]{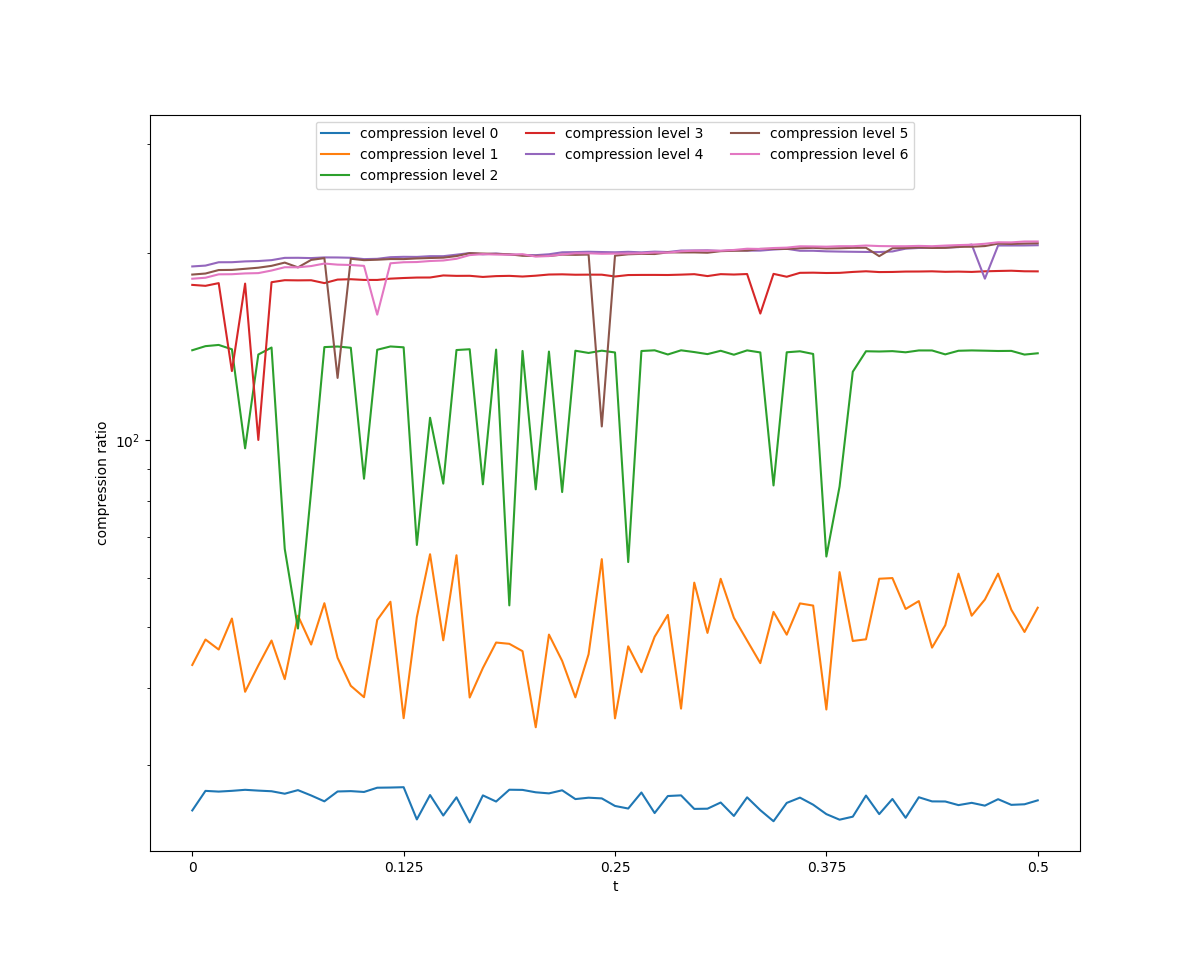}
        \caption{LZ4, chunk = 1MB}
        \label{fig:compression_over_time_4}
    \end{subfigure}
    \caption{This figure shows the compression ratios at each time step of the simulation with different lossless compression methods. The different curves (compression levels) represent the number of performed wavelet transforms.}
    \label{fig:compression_over_time}
\end{figure}

In this section, we show the compression ratios during the simulation with different lossless compression methods.
We test the following compression methods:
\begin{itemize}
    \item CSR (Figure \ref{fig:compression_over_time_1})
    \item LZ4 with a chunk size of 64 KB (Figure \ref{fig:compression_over_time_2})
    \item LZ4 with a chunk size of 256 KB (Figure \ref{fig:compression_over_time_3})
    \item LZ4 with a chunk size of 1 MB (Figure \ref{fig:compression_over_time_4})
\end{itemize}
We use a threshold value of 0.01.
For the numerical simulation, we use $\alpha=0.9$, $\beta=0.9$, and $CFL=0.45$.
The compression ratio is defined as the size of the uncompressed data divided by the size of the compressed data.

These figures show that the compression ratio is not constant during the simulation.
It tends to increase as the simulation progresses.
This trend seems to be exacerbated as the compression level increases.
This suggests that the data become more and more easily compressible as the simulation progresses.
This is likely because each time step allows for removing more and more details from the simulation.
An interesting observation is that the compression ratio tends to increase with the compression level.
Compression levels 5 and 6 are an exception to this trend but we can reasonably exclude them from the analysis.
Indeed, they add DWTs to the coarsest levels ($k=0$ and $k=1$), which have too few sampling points to perform a reasonable DWT.
These two additional levels can even create new non-null details, leading to a decrease in the compression ratio.

There are substantial differences between the CSR (Figure \ref{fig:compression_over_time_1}) and the LZ4 compression methods (Figures \ref{fig:compression_over_time_2}, \ref{fig:compression_over_time_3} and \ref{fig:compression_over_time_4}).
The CSR method produces smooth compression ratios over the simulation, while the LZ4 method produces more erratic compression ratios.
For the CSR method, this is explained by the fact that the compression ratio is directly dependent on the number of non-zero values which have no reason to brutally change from one time step to another.
For the LZ4 method, this is likely due to an inner mechanism of the LZ4 algorithm that makes it underperform in some time steps.
The obtained compression ratios are comparable for the CSR method (Figure \ref{fig:compression_over_time_1}) and the LZ4 method with a chunk size of 64 KB (Figure \ref{fig:compression_over_time_2}).
However, the LZ4 method with chunk sizes of 256 KB (Figure \ref{fig:compression_over_time_3}) and 1 MB (Figure \ref{fig:compression_over_time_4}) both outperform the CSR method.

In terms of compression ratio, LZ4 with a chunk size of 1 MB is the best method.
The CSR format is not designed to be a compression method.
It is, therefore, not surprising that it can be outperformed.
On the other hand, having a larger chunk size typically leads to increased compression ratios but at the cost of increased computation time.
Finding the right balance between compression ratio and computation time is a challenge that we will address in the future.

\subsubsection{Impact of the threshold value on the compression ratio} \label{sec:compression_ratio_threshold}

\begin{figure}[h]
    \centering
    \begin{subfigure}[b]{0.38\textwidth}
        \centering
        \includegraphics[width=\textwidth]{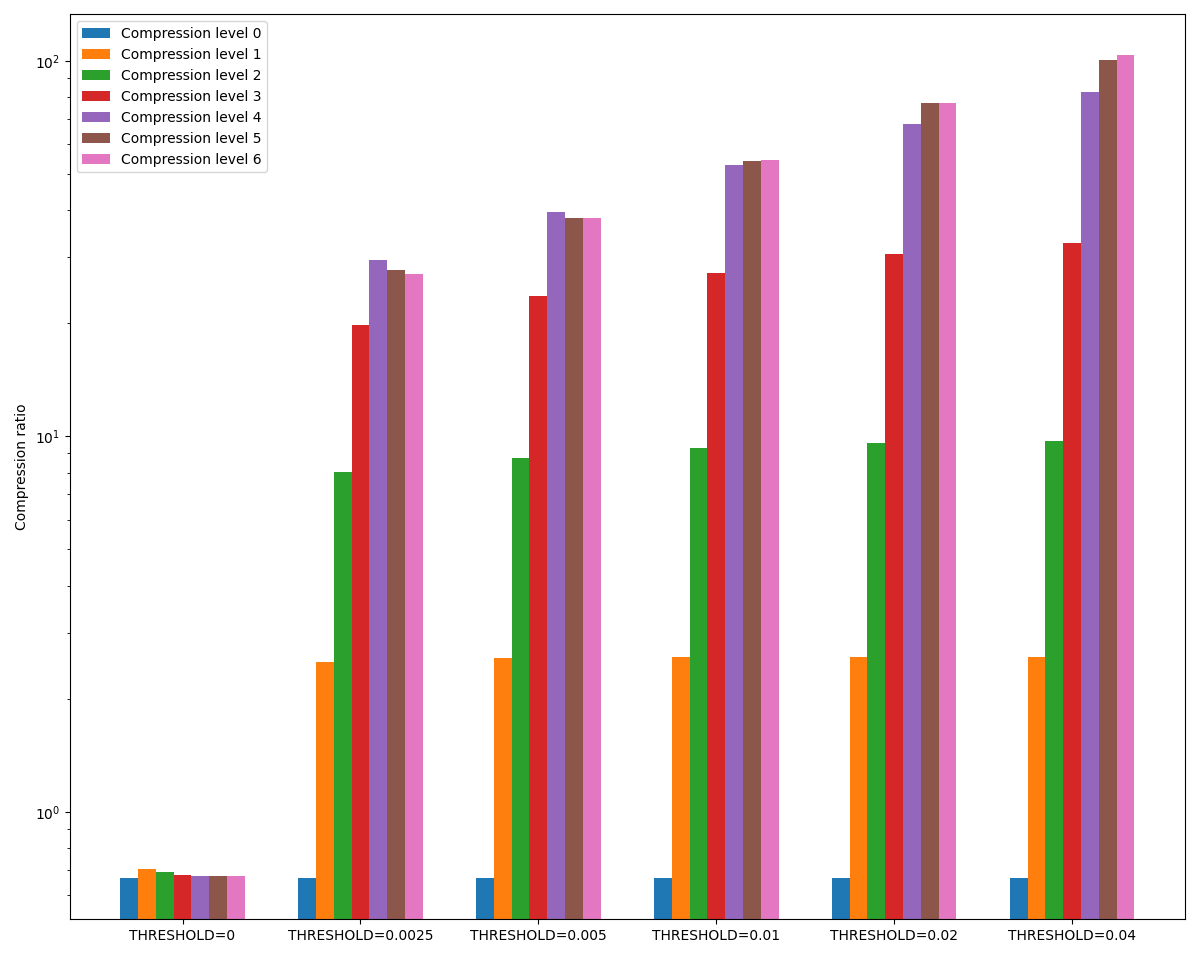}
        \caption{CSR}
        \label{fig:compression_threshold_1}
    \end{subfigure}
    \begin{subfigure}[b]{0.38\textwidth}
        \centering
        \includegraphics[width=\textwidth]{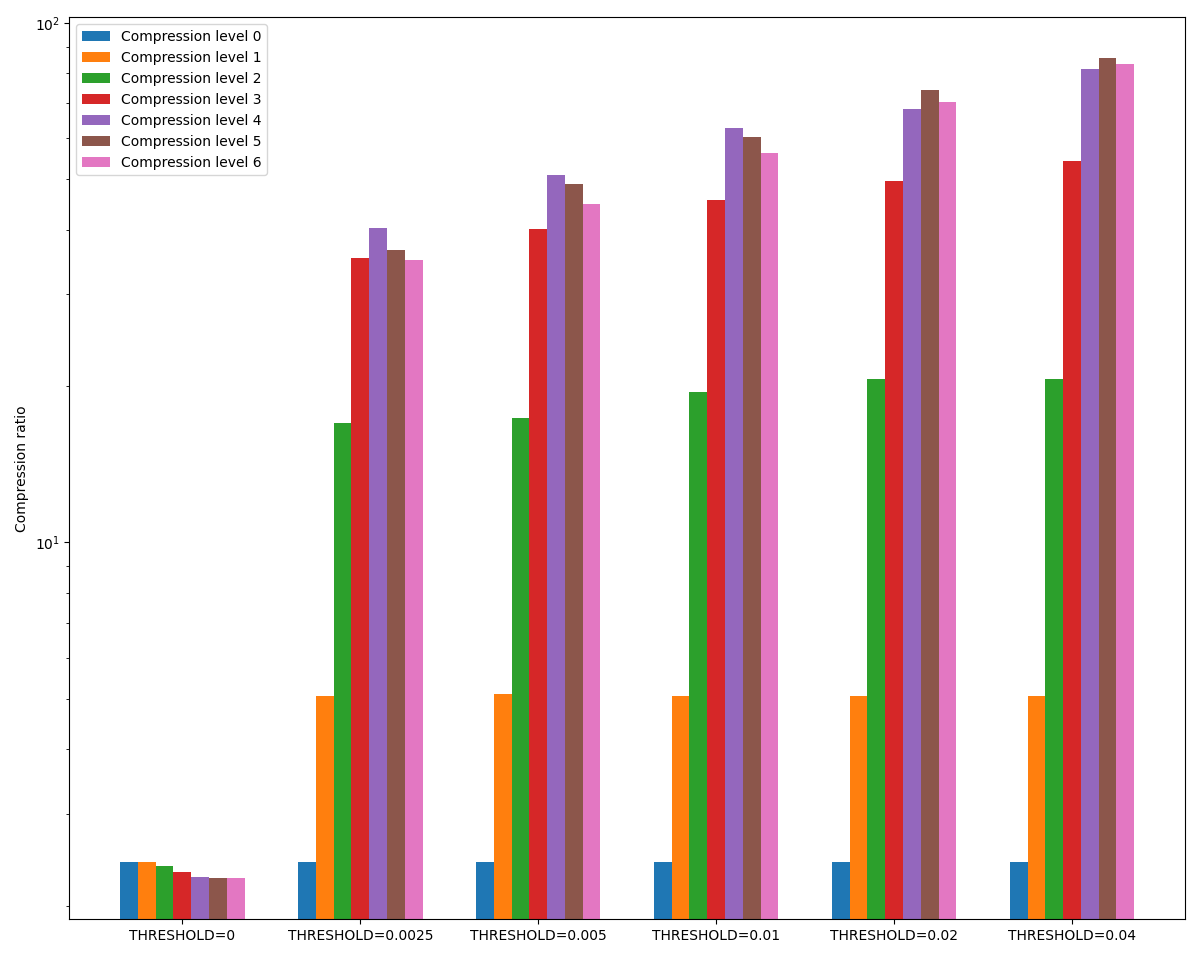}
        \caption{LZ4, chunk = 64KB}
        \label{fig:compression_threshold_2}
    \end{subfigure}
    \begin{subfigure}[b]{0.38\textwidth}
        \centering
        \includegraphics[width=\textwidth]{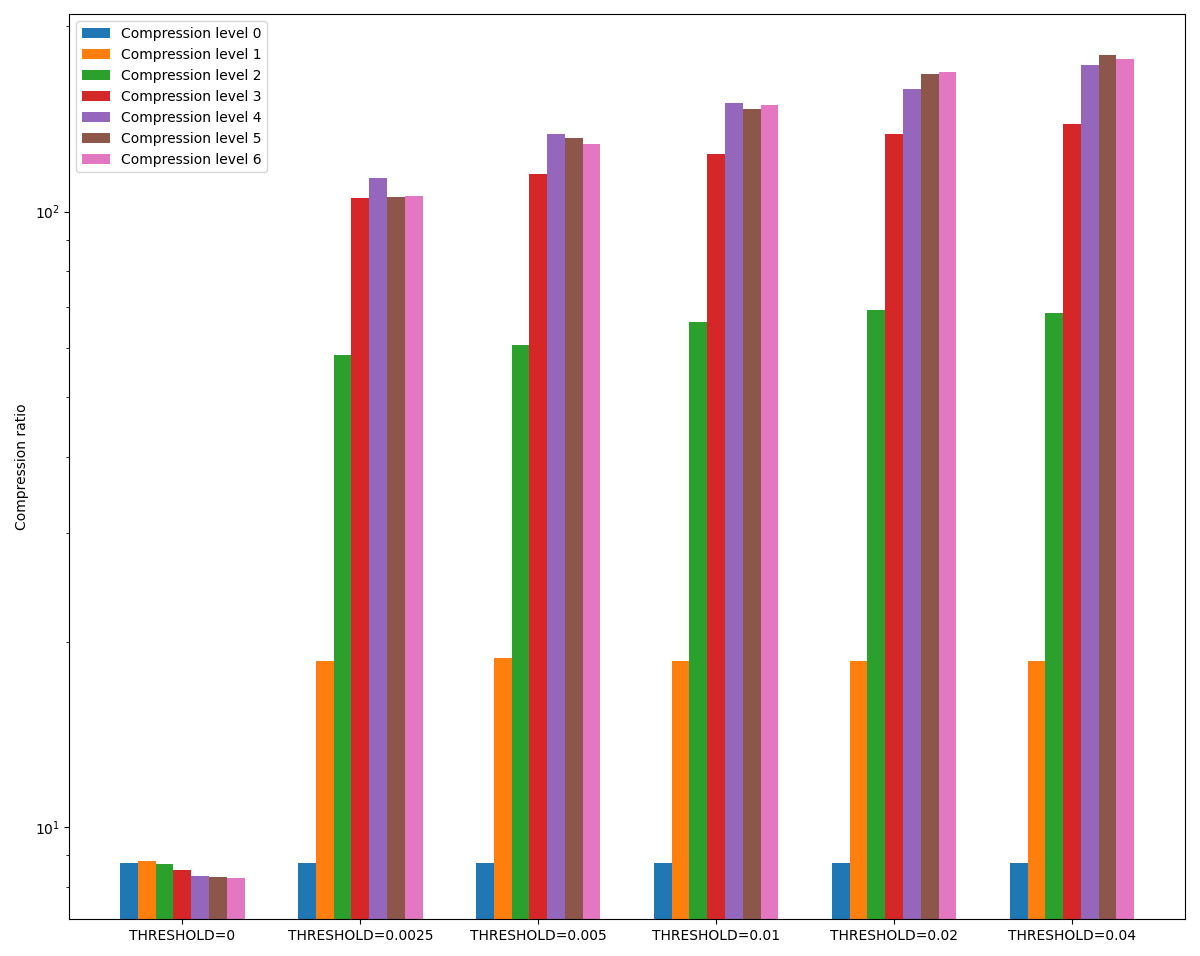}
        \caption{LZ4, chunk = 256KB}
        \label{fig:compression_threshold_3}
    \end{subfigure}
    \begin{subfigure}[b]{0.38\textwidth}
        \centering
        \includegraphics[width=\textwidth]{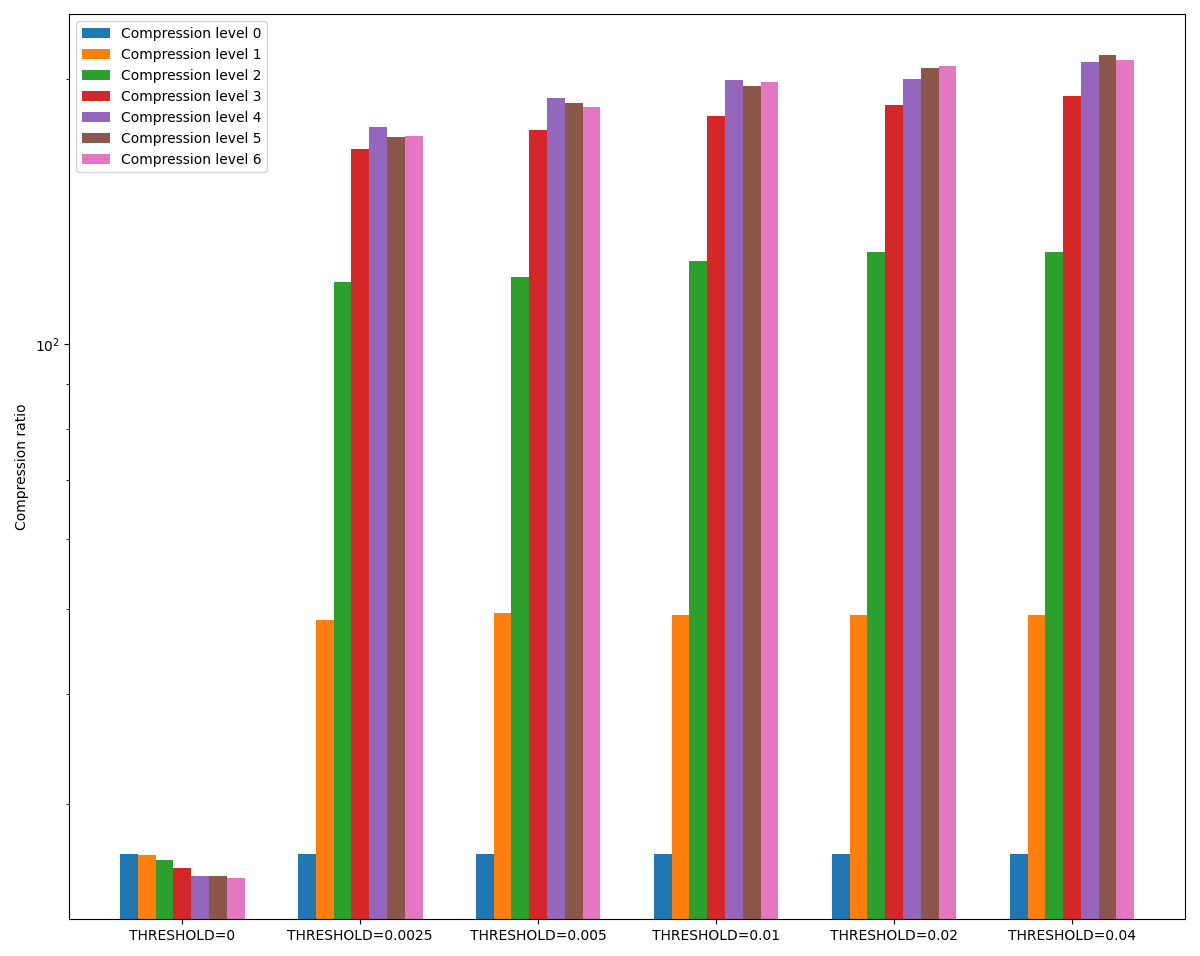}
        \caption{LZ4, chunk = 1MB}
        \label{fig:compression_threshold_4}
    \end{subfigure}
    \caption{This figure shows the average compression ratio for different threshold values and lossless compression methods. The compression level is the number of performed wavelet transforms.}
    \label{fig:compression_threshold}
\end{figure}

In this section, we aim to determine the impact of the threshold value on the compression ratio.
We keep the previous simulation parameters: $\alpha=0.9$, $\beta=0.9$, and $CFL=0.45$.
We test the following threshold values: 0, 0.0025, 0.005, 0.01, 0.02, and 0.04.
We show the results for 4 lossless compression methods: CSR (Figure \ref{fig:compression_threshold_1}), LZ4 with a chunk size of 64 KB (Figure \ref{fig:compression_threshold_2}), LZ4 with a chunk size of 256 KB (Figure \ref{fig:compression_threshold_3}) and LZ4 with a chunk size of 1 MB (Figure \ref{fig:compression_threshold_4}).

There is a general trend that the compression ratio tends to increase as the threshold value increases.
In general, increasing the compression level intensifies this trend.
For example, at compression level 1, the compression ratio always remains near-constant across non-zero threshold values while at compression level 4, the compression ratio always increases as the threshold value increases.
We can also reiterate the observation that up to a compression level of 4, the compression ratio tends to increase as the compression level increases.

The 4 methods show similar trends but are different in scale.
The CSR method (Figure \ref{fig:compression_threshold_1}) and the LZ4 method with a chunk size of 64 KB (Figure \ref{fig:compression_threshold_2}) are comparable in scale.
The LZ4 method with a chunk size of 256 KB (Figure \ref{fig:compression_threshold_3}) and with a chunk size of 1 MB (Figure \ref{fig:compression_threshold_4}) outperform the two previously mentioned methods.
With the 1 MB chunk size, the average effective compression ratio always reaches more than 100x if the threshold value and the compression level are greater than or equal to 0.0025 and 2, respectively.

The results, with a threshold value of 0, provide insight into the significance of the thresholding phase.
As observed in all four methods, the compression ratio significantly improves as the threshold value increases from 0 to 0.0025.
In the case of CSR compression, this outcome is expected, as the compression ratio is directly related to the number of non-zero values.
For LZ4 compression, thresholding the data increases the frequency of the "zero" symbol, resulting in a higher compression ratio.
However, it is worth noting that LZ4 can already achieve a high compression ratio without thresholding the data.
Indeed, with all chunk sizes, the compression ratio is always greater than 2x.
The 256 KB chunk size nearly achieves 9x average compression rate while the 1 MB chunk size nearly achieves more than 25x, both with no pretreatment (i.e. threshold value and compression level of 0).

This study suggests that setting a non-null threshold value is very important for achieving a high compression ratio.
It also appears that the first 3 compression levels (performed DWTs) are the most impactful in our case in terms of compression ratio.

\subsubsection{Quality of the simulation} \label{sec:quality_simulation}

\begin{figure}[h]
    \centering
    \includegraphics[width=0.5\textwidth]{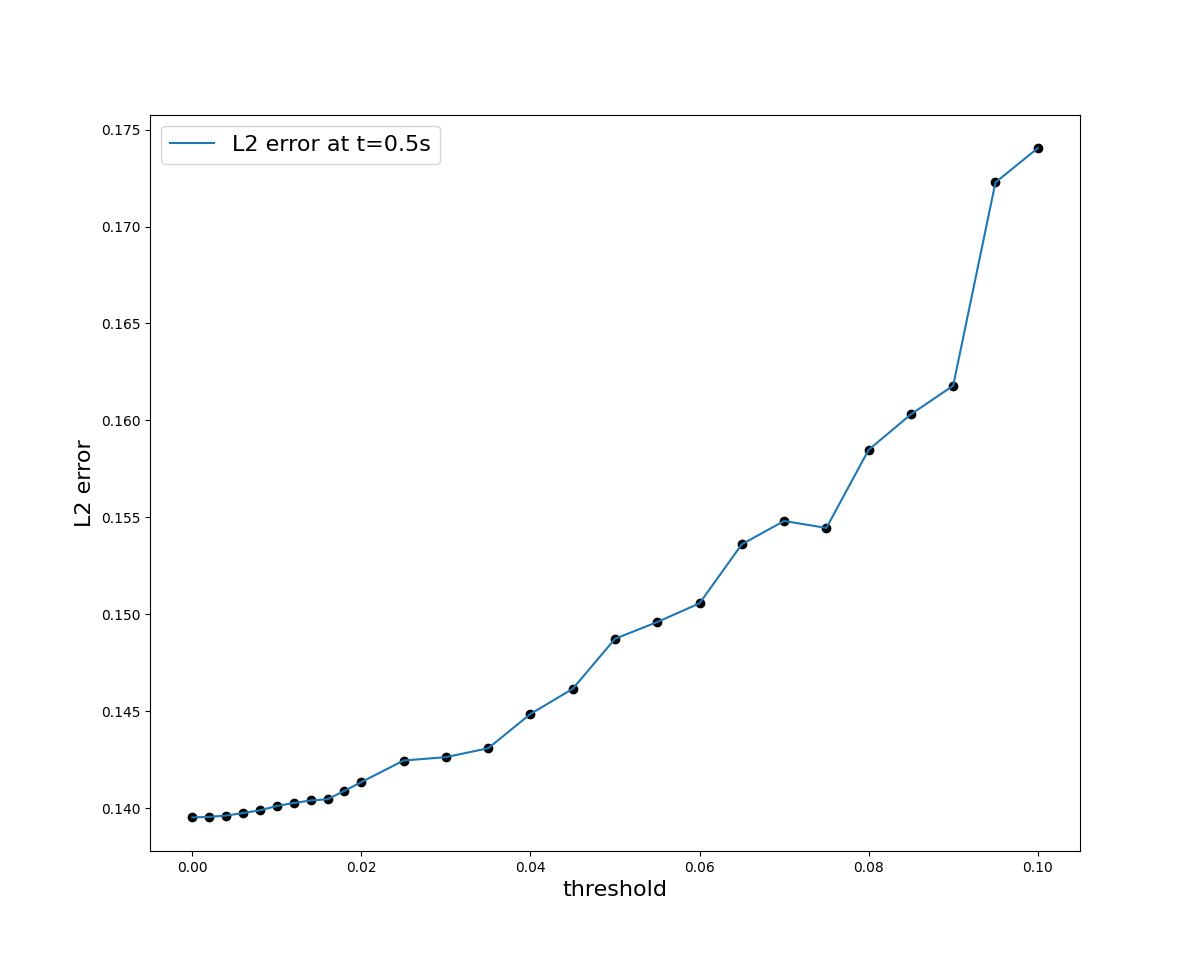}
    \caption{This figure shows the L2 error (formula \ref{eq:L2_error}) for different threshold values at compression level 4 and t=0.5s, with $\alpha=0.9$, $\beta=0.9$, and $CFL=0.45$.}
    \label{fig:L2_error}
\end{figure}

\begin{figure}[htbp]
    \centering
    \begin{subfigure}[b]{0.24\textwidth}
        \centering
        \includegraphics[width=\textwidth]{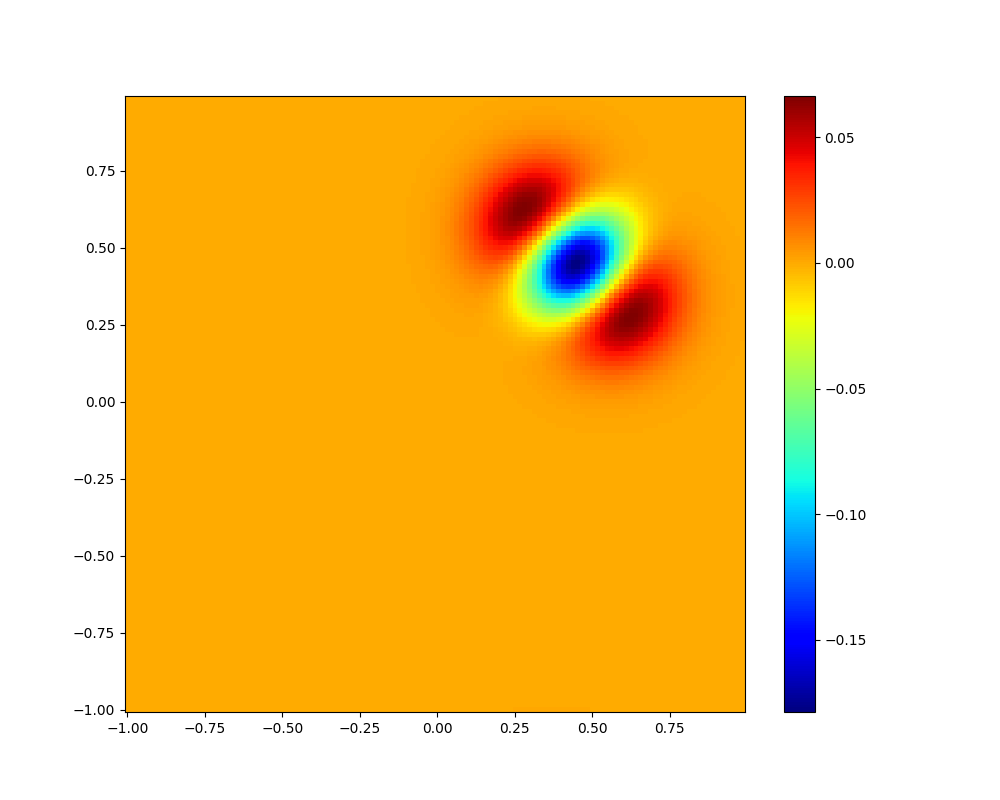}
        \caption{threshold=0}
        \label{fig:error_plot_1}
    \end{subfigure}
    \begin{subfigure}[b]{0.24\textwidth}
        \centering
        \includegraphics[width=\textwidth]{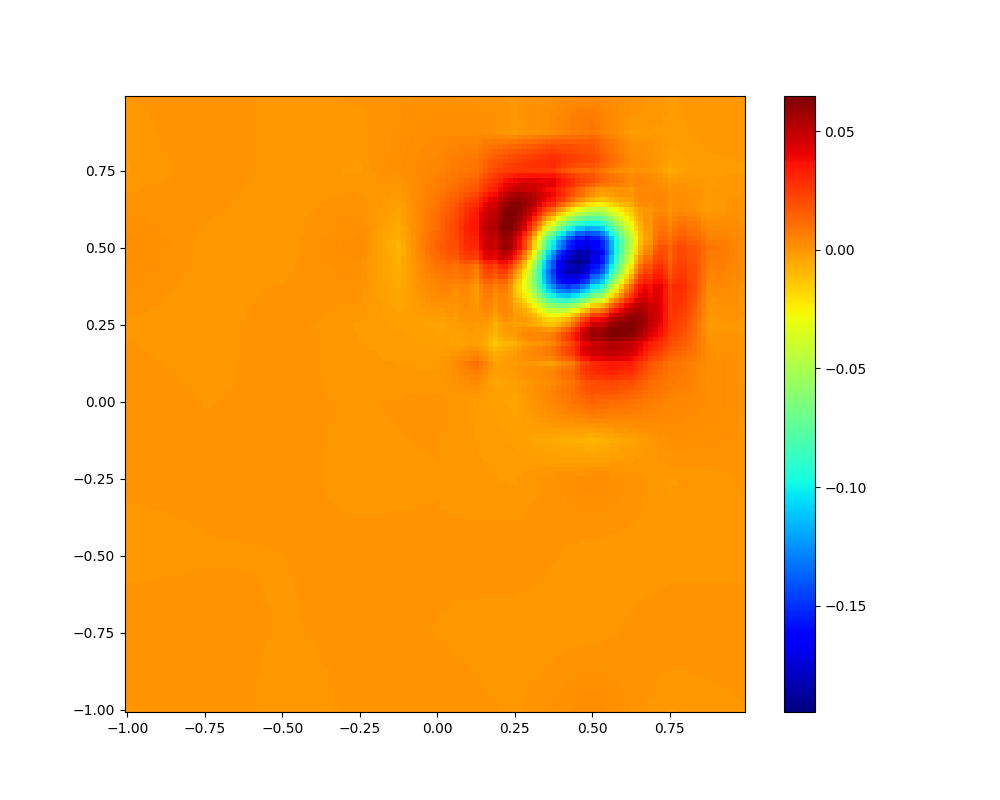}
        \caption{threshold=0.01}
        \label{fig:error_plot_2}
    \end{subfigure}
    \begin{subfigure}[b]{0.24\textwidth}
        \centering
        \includegraphics[width=\textwidth]{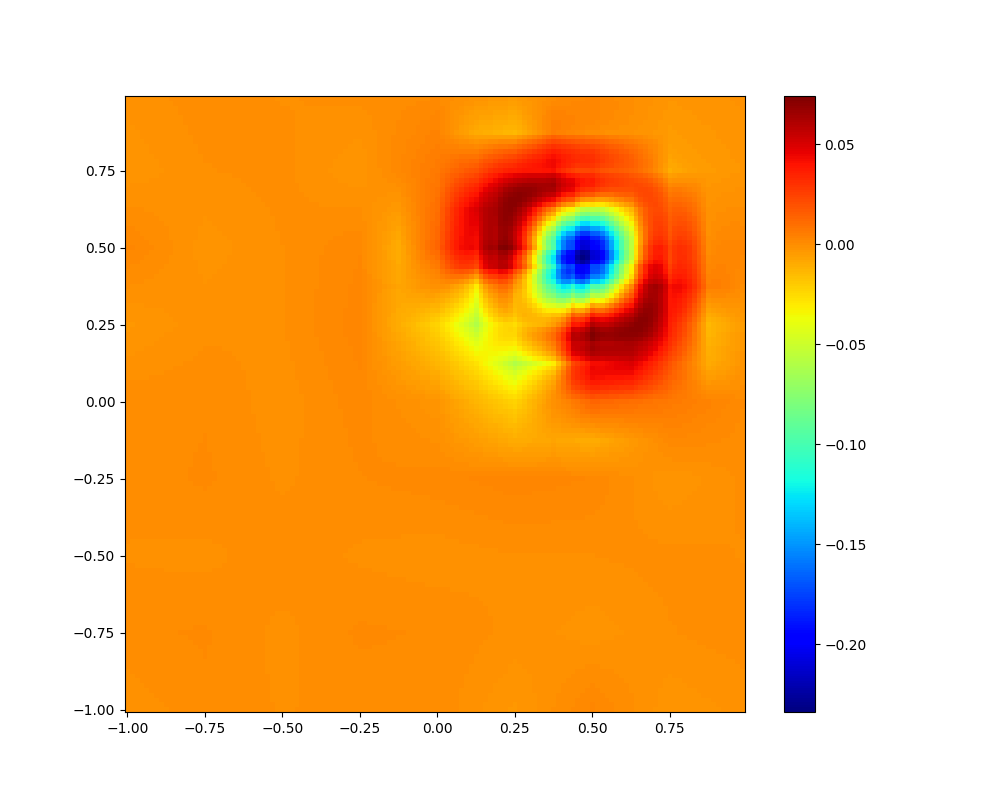}
        \caption{threshold=0.02}
        \label{fig:error_plot_3}
    \end{subfigure}
    \begin{subfigure}[b]{0.24\textwidth}
        \centering
        \includegraphics[width=\textwidth]{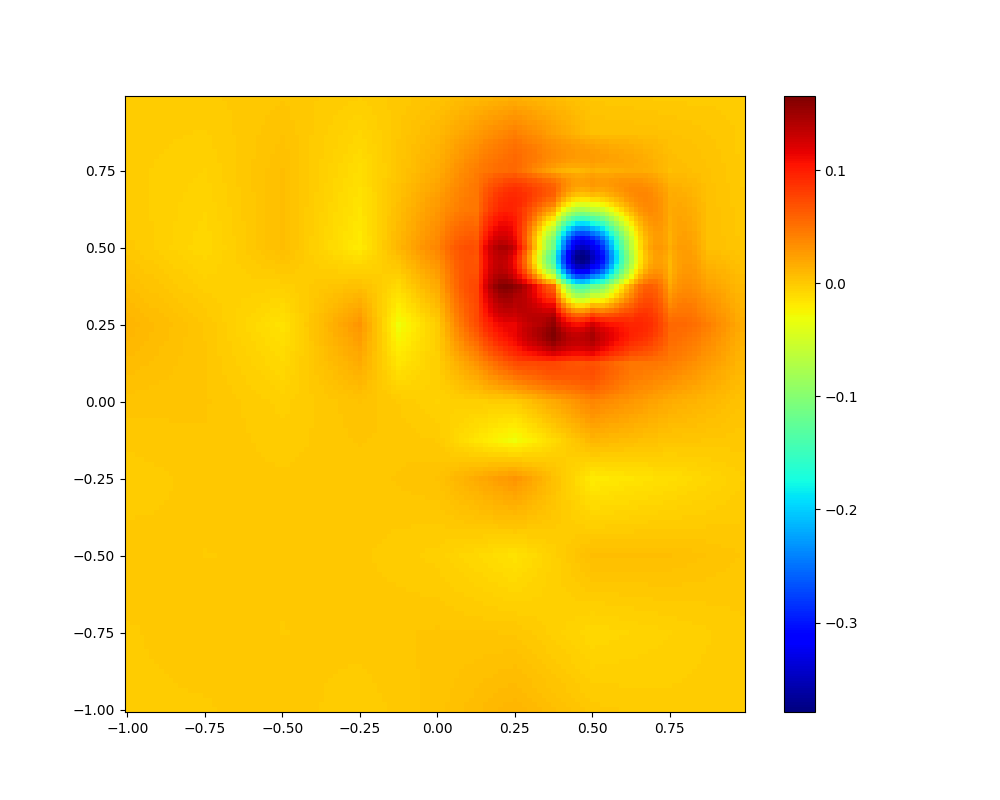}
        \caption{threshold=0.05}
        \label{fig:error_plot_4}
    \end{subfigure}
    \caption{These plots show how the error is distributed over the domain at $t=0.5s$ and with a compression level of 4 for different threshold values.}
    \label{fig:error_plot}
\end{figure}

We evaluate the quality of the simulation by comparing the results of the simulation with the exact solution.
The L2 error (formula \ref{eq:L2_error}) is measured at the end of the simulation ($t=0.5s$).
Figure \ref{fig:L2_error} shows the L2 error for different threshold values.

The first value at threshold=0 provides the L2 error at the end of the simulation with no thresholding, i.e. with no loss due to compression.
The other values show the impact of the chosen threshold value on the L2 error.
We can see that the L2 error tends to increase as the threshold value increases.
This is expected, as each time a value is thresholded, it is replaced by 0, which leads to a greater error in the reconstruction of the signal.

Figure \ref{fig:error_plot} shows how the error is distributed over the domain at $t=0.5s$ for different threshold values.
The plotted error is the difference between the simulation result and the exact solution: $error(i,j) = f\_sim(i,j) - f\_exact(i,j)$.
We can see the location of the simulation error when there is no loss due to compression (i.e. threshold=0)
In the original simulation (threshold=0), the errors mainly appear in the center of the Gaussian structure (in blue) and in the two directions perpendicular to its ($\alpha$,$\beta$) speed (in red).
Visually, the Gaussian structure is stretched from either side of its speed.
As the threshold value increases, the error becomes more and more distorted compared to that of the original simulation.

This study quantifies the impact of the threshold value on the quality of the simulation.
In our simulation, a threshold value between 0 and 0.05 has little visual impact on the simulation results.
The acceptable error level will vary depending on the specific application.
A major advantage of our approach is that the induced error can be controlled by the user, through the threshold value.

\subsubsection{Computational cost} \label{sec:computational_cost}

It is important to note that the compression and decompression kernels presented in this study are not intended to reach the highest level of optimisation.
Rather, the purpose of evaluating the computational cost is to provide a reference point for the reader and to give an estimate of the potential impact of our method on a real simulation.
As such, the results should be considered as an indicative measure of the performance and not as an absolute representation of the optimisation level achievable with further fine-tuning.

\begin{figure}[htbp]
    \centering
    \begin{subfigure}[b]{0.32\textwidth}
        \centering
        \includegraphics[width=\textwidth]{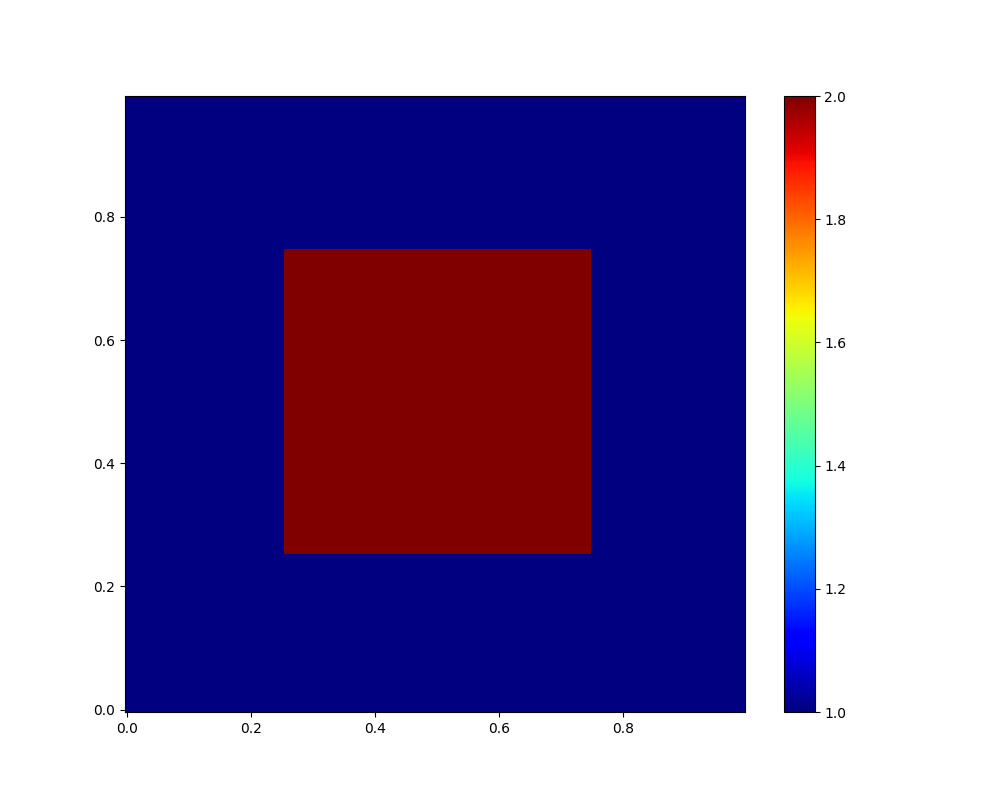}
        \caption{t=0s, degraded}
        \label{fig:stvenant_t0}
    \end{subfigure}
    \begin{subfigure}[b]{0.32\textwidth}
        \centering
        \includegraphics[width=\textwidth]{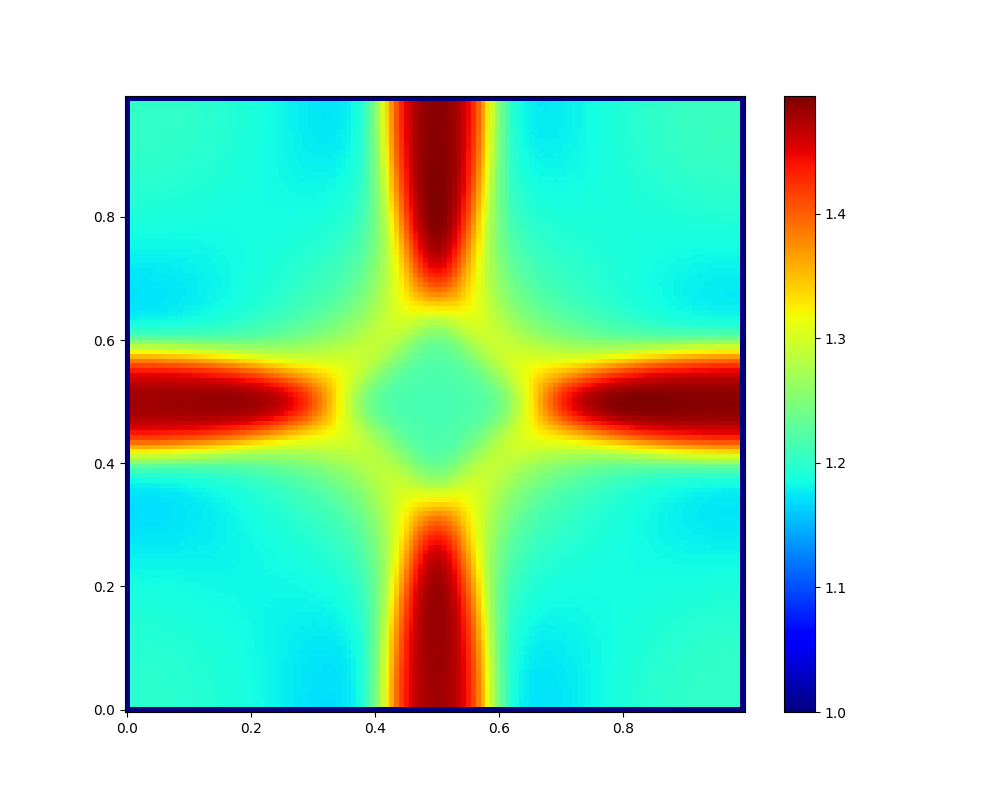}
        \caption{t=0.5s, original}
        \label{fig:stvenant_t0.5_original}
    \end{subfigure}
    \begin{subfigure}[b]{0.32\textwidth}
        \centering
        \includegraphics[width=\textwidth]{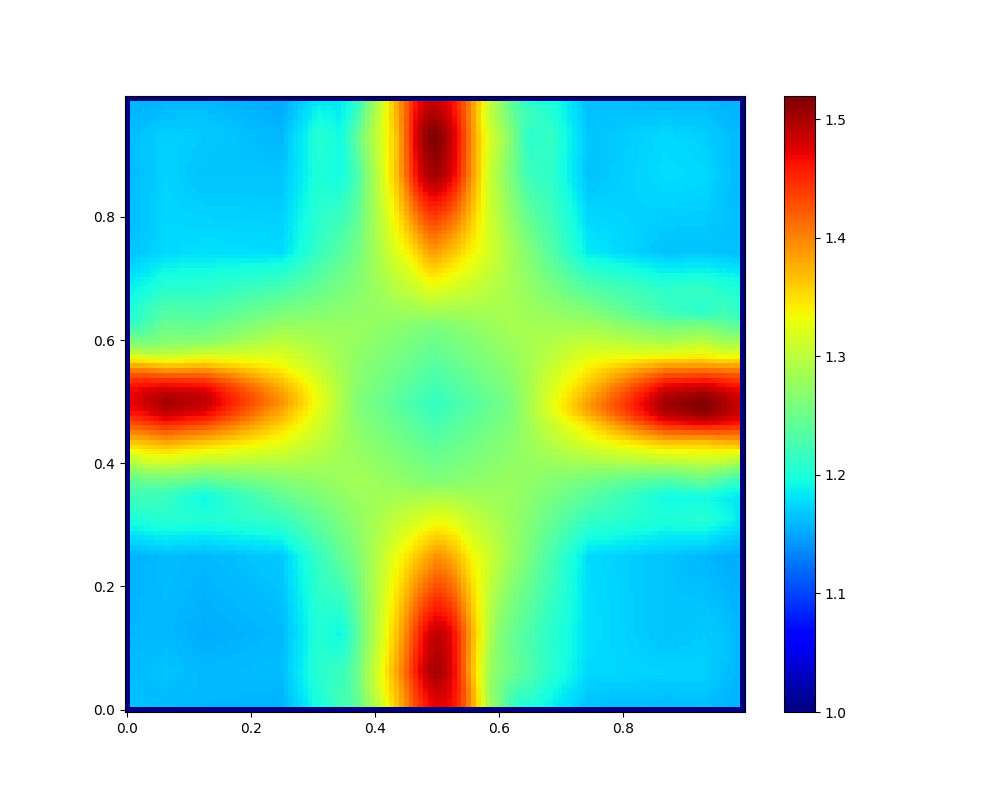}
        \caption{t=0.5s, degraded}
        \label{fig:stvenant_t0.5_degraded}
    \end{subfigure}
    \begin{subfigure}[b]{0.32\textwidth}
        \centering
        \includegraphics[width=\textwidth]{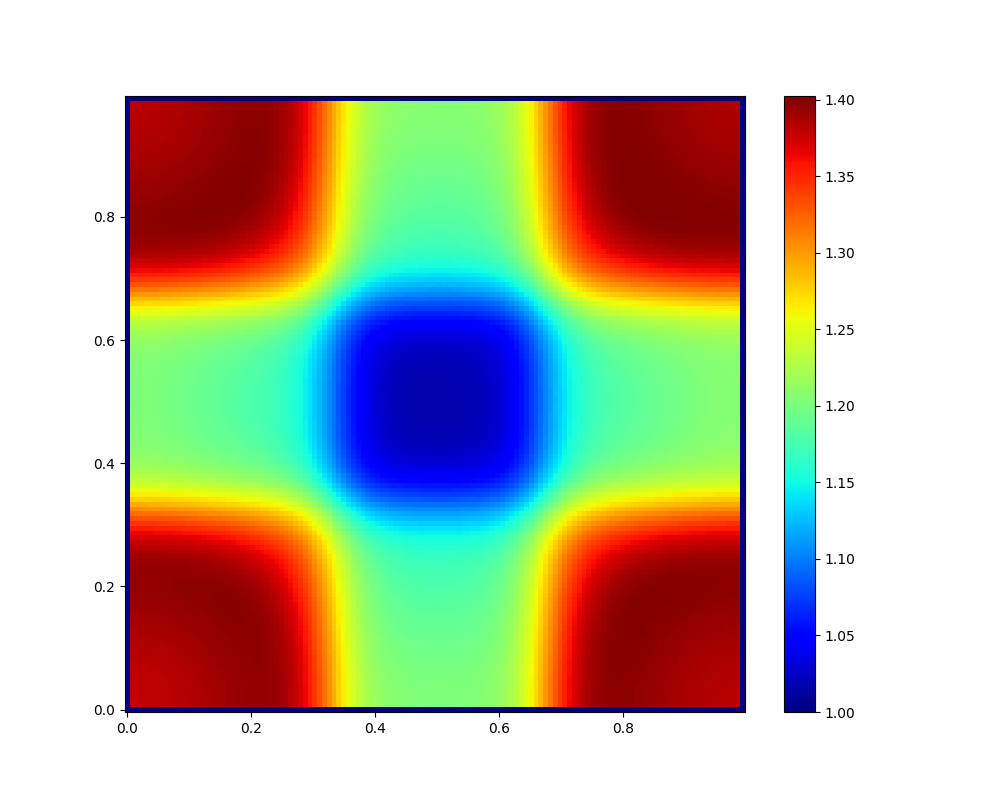}
        \caption{t=1s, original}
        \label{fig:stvenant_t1_original}
    \end{subfigure}
    \begin{subfigure}[b]{0.32\textwidth}
        \centering
        \includegraphics[width=\textwidth]{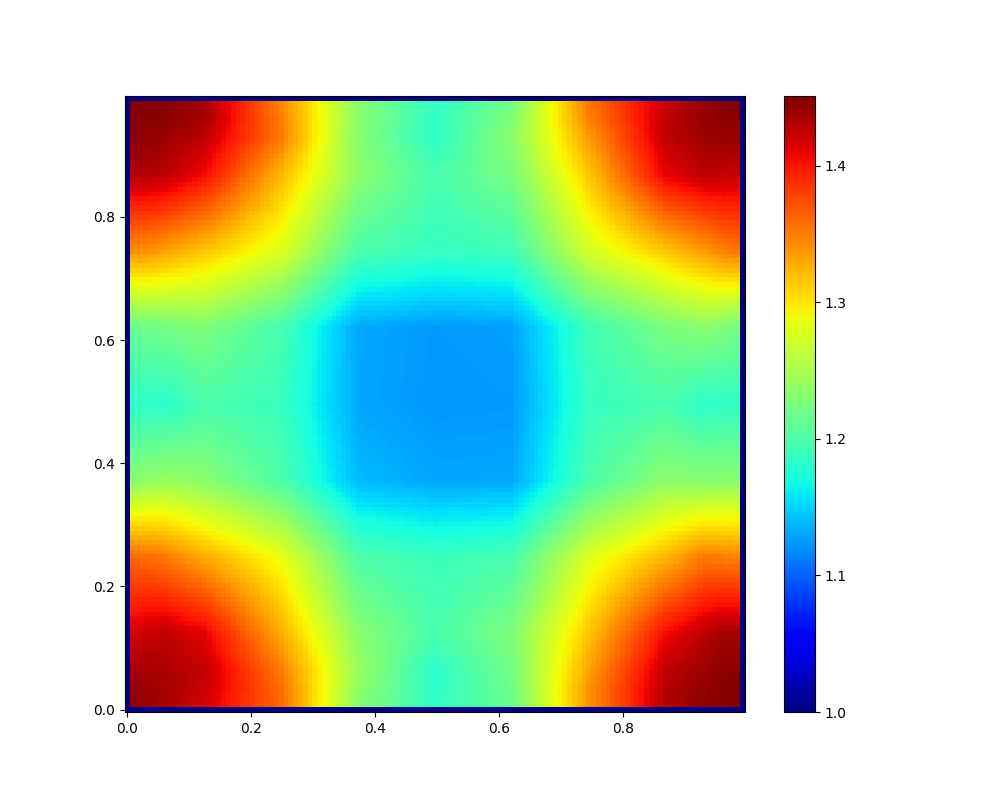}
        \caption{t=1s, degraded}
        \label{fig:stvenant_t1_degraded}
    \end{subfigure}
    \caption{These plots show the results of the simulation with no compression (original) and with the DWT (degraded). The threshold value is 0.0005.}
    \label{fig:stvenant}
\end{figure}

To evaluate the computational cost of the compression pipeline on a more computationally intensive  simulation, we use a Godunov scheme to solve a shallow water model. The shallow water model is defined by
\begin{equation}
    W=\left(\begin{array}{c}
        h\\
        hu\\
        hv
        \end{array}\right),\quad Q(W,N)=\left(\begin{array}{c}
        h(uN^{x}+vN^{y})\\
        hu(uN^{x}+vN^{y})+\frac{1}{2}gh^{2}N^{x}\\
        hv(uN^{x}+vN^{y})+\frac{1}{2}gh^{2}N^{y}
        \end{array}\right),
\end{equation}
where $h$ is the water level, $(u,v)$ the horizontal velocity vector and $g=9.81\text{m/s}^2$ the gravitational acceleration. In the following simulations, we use the Godunov numerical flux, based on exact Riemann solvers (we refer, for instance, to\cite{leveque2002finite} for the details).
At $t=0$, there is a $0.5$-meter square where the water level is $h=2$ meters high and the rest of the domain is at $h=1$ meter (Figure \ref{fig:stvenant_t0}).
We run the simulation on a 129x129 grid and perform 38245 time steps, which correspond to a simulation time of 10 seconds.
We use a compression level (number of DWTs) of 4 and a threshold value of 0.0005.
We only show the results for the CSR lossless compression because the LZ4 compression provided poor compression ratios due to the method we used to implement fast wavelets (more details on this in the following).
The lossless CSR compression is implemented using the \texttt{cuSPARSE} library.

\begin{table}[htbp]
    \centering
    \begin{tabular}{|l|c|c|c|}
        \hline
        \multirow{2}{*}{Kernel} & \multicolumn{3}{c|}{Execution time} \\
        \cline{2-4}
        & no compression & only wavelets & wavelets + CSR \\
        \hline
        total GPU time & 95.6735s & 110.754s & 114.126s \\
        time\_step & 95.6735s & 97.1642s & 97.2008s \\
        wavelet\_x\_compress & $\emptyset$ & 2.79200s & 2.41154s \\
        wavelet\_step\_y\_compress\_samples & $\emptyset$ & 2.33168s & 2.25091s \\
        wavelet\_step\_y\_compress\_details & $\emptyset$ & 2.30558s & 2.40358s \\
        wavelet\_x\_decompress & $\emptyset$ & 1.56730s & 1.08666s \\
        wavelet\_step\_y\_decompress\_samples & $\emptyset$ & 2.30878s & 2.23454s \\
        wavelet\_step\_y\_decompress\_details & $\emptyset$ & 2.28504s & 1.55381s \\
        dense\_to\_csr & $\emptyset$ & $\emptyset$ & 1.08666s \\
        csr\_to\_dense & $\emptyset$ & $\emptyset$ & 295.01ms \\
        cusparseParseDenseByRows\_kernel & $\emptyset$ & $\emptyset$ & 620.01ms \\
        other\_cusparse\_kernels & $\emptyset$ & $\emptyset$ & 954.9ms \\
        \hline
        overhead & 0\% & +13.99\% & +17.41\% \\
        \hline
    \end{tabular}
    \caption{Execution times of the different kernels for the Godunov simulation.}
    \label{tab:execution_times}
\end{table}

Table \ref{tab:execution_times} shows the execution times of the different kernels for the Godunov simulation.
Each row provides the total time spent in the kernel.
The \texttt{total GPU time} row shows the total time spent performing GPU computations.
The \texttt{time\_step} row corresponds to the execution of a time step.
The \texttt{wavelet\_...} kernels correspond to the DWT kernels.
These include wavelet compression and decompression (inverse DWT) operations along the X and Y axes.
The wavelet\_y kernels are split into two parts and only perform one step of the DWT.
The wavelet\_x kernels perform the whole DWT along the X-axis in one kernel.
Additionally, the \texttt{dense\_to\_csr} and \texttt{csr\_to\_dense} kernels refer to the conversion between dense and compressed sparse row representations.
The \texttt{cusparseParseDenseByRows\_kernel} is an internal cuSPARSE kernel.
The remaining internal cuSPARSE kernels are grouped in the \texttt{other\_cusparse\_kernels} row.
Finally, the \texttt{overhead} row shows the percentage of overhead introduced by the compression method.

The LZ4 compression has not been tested on this simulation because it provides poor compression ratios.
The reason for this is that the optimised DWT kernels used in this study are different from the ones used in the previous sections.
In the previous sections, the samples are grouped in a corner of the matrix and the details are stored in the remaining cells.
This is convenient for the LZ4 compression because the details tend to be contiguous in memory, leading to a higher chance of finding matches in this area.
In the optimised DWT kernels, the samples are distributed evenly in the matrix and the details are stored in the remaining cells.
This is a worse scenario for the LZ4 compression because the dispersed samples make it harder to find matches.

These results let us showcase the benefits of the compression pipeline.
We compute an average compression rate of \textbf{x73.51} during the first 1 second of the simulation.
The overhead introduced by the compression pipeline is less than 20\% in this simulation.
This overhead is acceptable, especially considering that the compression ratio is extremely high.
On the other hand, the quality of the simulation can be assessed by comparing the results of the simulation with no compression (original) and with the DWT (degraded) in Figure \ref{fig:stvenant}.
We can see that the general outline of the simulation is preserved, even though the details are lost.
More details can be kept by decreasing the threshold value.
Overall, this performance study demonstrates the possibility of using our runtime compression pipeline to compress large simulations.
The current implementation does not let the user exceed the GPU memory capacity because the whole grid is decompressed at the same time.
However, this last limitation can be overcome by having only a few subgrids decompressed at a time.
In future work, we plan to use the task-based programming model to implement this idea efficiently.

\section*{Conclusions} \label{sec:conclusion}

In this work, we have adapted a compression scheme that combines a discrete wavelet transform with a lossless compression algorithm for optimising the memory management of numerical simulations on regular grids.
This algorithm is designed to allow a controlled loss of information while ensuring conservation of the global mass throughout the simulation.
We have shown that in a 128x128 grid, 2-dimensional compression can achieve a ratio of approximately 200x on a simple transport equation.
On a more complex simulation based on the shallow water equations, we have shown that the compression ratio can reach more than 70x.
This is substantial and can help to fit large simulations in the GPU memory.
With this approach, simulation schemes that are traditionally thought to be unadapted to GPU computing because of their high memory requirements could become feasible without changing the FV/LBM kernels.
The downside of this approach is that the compression is lossy.
However, we have shown that only a small loss of information is necessary to achieve a high compression ratio.
In addition, this loss can be controlled by the user, with the threshold value.
We have shown that on a shallow-water simulation, the compression/decompression pipeline only slows down the simulation by less than 20\%.
This slowdown is the only additional cost of our method and it can reduce the memory requirements by multiple orders of magnitude.
To be able to run an execution where the simulation grid actually exceeds the GPU memory, only a few subgrids should be decompressed in global memory at a time.
In our current implementation, this is not possible because the synchronisation between the subgrids assumes that the neighbours are also decompressed.
This is why our next work will consist of running a large-scale simulation on a GPU with a memory that is not large enough to fit the whole simulation grid.
Additionally, we plan to improve the efficiency of our compression pipeline to take full advantage of the GPU architecture.

\bibliographystyle{abbrv}
\bibliography{refs.bib} 

\end{document}